\documentclass[]{pasj01}

\Received{2020/07/30}
\Accepted{2021/03/30}
\Published{2021/05/01}
\jyear{2021}
\doi{psab028}
\endpage{676}
\volume{73}
\Issue{3}
 
\usepackage{xspace, bm, natbib}
\bibpunct[:]{(}{)}{,}{a}{}{,}
\newcommand{\ourcode}{GPLUM\xspace}
 
\begin{document} 

\title{ 
Particle--particle particle--tree Code for Planetary System Formation with Individual Cut-off Method: \ourcode
 }

\author{Yota \textsc{Ishigaki}\altaffilmark{1,2}%
\thanks{7-3-1, Hongo, Bunkyo-ku, Tokyo, 113-8654, Japan}}
\altaffiltext{1}{Department of Earth and Planetary Sciences, Graduate School of Scienece, University of Tokyo}
\altaffiltext{2}{Department of Solar System Science,
Institute of Space and Astronautical Science, Japan Aerospace Exploration Agency}
\email{y.ishigaki@stp.isas.jaxa.jp}

\author{Junko \textsc{Kominami}}\altaffilmark{3}%
\altaffiltext{3}{Earth-Life Science Institute, Tokyo Institute of Technology, 2-12-1-IE-1 Ookayama, Meguro-ku, Tokyo 152-8550, Japan}

\author{Junichiro \textsc{Makino}\altaffilmark{4,5}%
\thanks{1-1, Rokkodai-cho, Nada-ku, Kobe, Hyogo, 657-8501, Japan}}
\altaffiltext{4}{Department of Planetology, Graduate School of Science, Kobe University}
\altaffiltext{5}{RIKEN Center for computational Science}

\author{Masaki \textsc{Fujimoto}\altaffilmark{2}%
\thanks{3-1-1, Yoshinodai, Chuo-ku, Sagamihara, Kanagawa, 252-5210, Japan}}

\author{Masaki \textsc{Iwasawa}\altaffilmark{6}%
\thanks{14-4, Nishiikuma-cho, Matsue, Shimane, 690-8518}}
\altaffiltext{6}{National Institute of Technology, Matsue College}


\maketitle

\begin{abstract}
In a standard theory of the formation of the planets in our Solar System,
terrestrial planets and cores of gas giants are formed through accretion of kilometer-sized objects (planetesimals) in a protoplanetary disk. 
Gravitational $N$-body simulations of a disk system made up of numerous planetesimals are the most direct way to study the accretion process.
However, the use of $N$-body simulations has been limited to idealized models (e.g. perfect accretion) and/or narrow spatial ranges in the radial direction, due to the limited number of simulation runs and particles available.
We have developed new $N$-body simulation code equipped with a particle--particle particle--tree (${\rm P^3T}$) scheme for studying the planetary system formation process: \ourcode.
For each particle, \ourcode uses the fourth-order Hermite scheme to calculate gravitational interactions with particles within 
cut-off radii and the Barnes--Hut tree scheme for particles outside the cut-off radii. 
In existing implementations, ${\rm P^3T}$ schemes use the same cut-off radius for all particles, making a simulation become slower when the mass range of the planetesimal population becomes wider.
We have solved this problem by allowing each particle to have an appropriate cut-off radius depending on its mass, its distance from the central star, and the local velocity dispersion of planetesimals. 
In addition to achieving a significant speed-up, we have also improved the scalability of the code to reach a good strong-scaling performance up to 1024 cores in the case of $N=10^6$.
\ourcode is freely available from \textcolor{blue}{https://github.com/YotaIshigaki/GPLUM} with MIT license.
\end{abstract}

\section{Introduction}

In the standard model of the planetary system formation, planets are considered to form from planetesimals. While how (and whether) planetesimals form in a protoplanetary disk is still under debate, a standard scenario assumes that a protoplanetary disk was initially filled with numerous planetesimals and the evolution of the system through gravitational interactions among planetesimals is considered to realize the scenario described below, 

Planetesimals coagulate and form runaway bodies within about $10^5\, {\rm yr}$ \citep[e.g.][]{Kokubo1996}.  This is called the runaway phase, and is followed by the oligarchic phase\citep[e.g.][]{Kokubo1998} in which the runaway bodies grow until they reach the isolation mass\citep[e.g.,][]{Kokubo2002}.  
The isolation mass is the mass at which the runaway bodies consume the planetesimals in their neighborhood. The separation distance between runaway bodies is within $5$--$10$ Hill radii. 
The isolation mass is $\sim\! M_{\rm Mars}$ in the terrestrial planet region.  In the gas giant and ice giant regions, isolation mass reaches several times $\sim\! M_\oplus$.  Once the mass of a runaway body reaches the critical value, which is about 10 times the Earth mass, runaway gas accretion starts to form a giant planet, with a runaway body constituting the core of the planet\citep{Mizuno1978,Mizuno1980,ikoma1998formation}.

One limitation of this scenario is that it is based on a rather limited set of $N$-body simulations, with a low mass resolution and a narrow radial range. For example, Kokubo and Ida (2002) used particles with a minimum mass of $10^{23}\, {\rm g}$ (corresponding to a $\sim$100 km-sized planetesimal), and a radial range of $0.5$--$1.5\, {\rm au}$. 

These limitations imply that our current understanding of the planetary system formation process is based on ``local'' physical models, in which we assume that the radial migration of seed planetesimals or protoplanets does not affect the formation process significantly. This assumption of locality is, however, clearly insufficient, especially when we recognize that some exoplanetary systems are clearly shaped by migration effects. Protoplanets and planets can
shift their radial positions via at least three main mechanisms: Type I migration, planetesimal-driven migration and interactions between
planets. In order to model the planetesimal-driven migration of protoplanets, a mass resolution much higher than in previous $N$-body simulations is necessary \citep{Minton2014}. When a protoplanet with a mass larger than typical planetesimals is formed, neighboring planetesimals are disturbed gravitationally. If the planetesimal population is not constructed with a good enough resolution, as in previous studies, the perturbation becomes random, making systematic radial migration artificially absent.To study the effects of migration, a simulation needs to involve a wide spatial range in the radial direction as well. 

Both a high resolution and a large spatial extent are, however, computationally demanding. Almost all previous long-term (covering more than $10^4\, {\rm yr}$) simulations have been performed with high-accuracy direct methods\citep[][]{Kokubo1996,Kokubo1998}, and some with acceleration by GRAPE hardware \citep[e.g.][]{sugimoto1990special,makino1993harp, Makino2003}. Since the calculation cost of the direct method is $O(N^2)$, where $N$ is the number of particles, using more than $10^6$
particles to realize a high resolution or a large spatial extent is impractical even with Japanese K computer or its successor, the Fugaku supercomputer, unless a new calculation scheme is introduced.

Oshino, Funato, and Makino (2011) developed a numerical algorithm which combines
the fast Barnes--Hut tree method \citep{Barnes1986} and an accurate and efficient individual time step Hermite integrator \citep{Aarseth1962, Makino1991} through Hamiltonian
splitting. This algorithm is called the particle--particle particle--tree, or ${\rm P^3T}$, algorithm. 

\citet{Iwasawa2017} reported the implementation and performance of a parallel ${\rm P^3T}$ algorithm developed using the FDPS framework \citep{Iwasawa2016}. 
FDPS is a general-purpose, high-performance library for particle simulations.
\citet{Iwasawa2017} showed that ${\rm P^3T}$ algorithm shows high performance even in simulations with large number of particle ($N=10^6$) and wide radial range ($1$--$11\, {\rm au}$).
Its performance scales reasonably well for up to 512 cores for one million particles, but with one limitation. The cut-off length used to split the Hamiltonian is fixed and shared by all the particles. Thus,
when the mass range of the particles becomes large through the runaway growth, the calculation efficiency is reduced substantially. If we take into account the collisional disruption of planetesimals, this problem becomes even more serious.

In this paper, we report on the implementation and performance of the GPLUM code for large-scale $N$-body simulation of planetary system formation, based on a parallel ${\rm P^3T}$ algorithm with individual mass-dependent cut-off length. We will show that the use of individual mass-dependent cut-off can speed up a calculation by a factor of $3$--$40$.

In section 2 we describe the implementation of GPLUM.  Section 3 is devoted to performance evaluation, and section 4 provides discussion and conclusion.

\section{Numerical method}
If the fourth order Hermite scheme with individual time step \citep{Aarseth1962} is used, the order of the calculation cost becomes $O(N^2)$, where $N$ is the number of particles.  
Hence, the cost of calculations increases significantly as $N$ increases. 
Simulations using parallelized code on their Japanese K computer \citep[e.g.][]{Kominami2016} could treat up to several times $10^5$ particles.  They used the Ninja algorithm \citep{Nitadori2006} for parallelization of the Hermite scheme.  
On the other hand, in order to increase the number of particles that can be treated, the tree method \citep{Barnes1986}
has been used at the expense of numerical accuracy.  
In order to improve both accuracy and speed, a scheme which combines the Barnes--Hut tree scheme and the high-order Hermite scheme (${\rm P}^3{\rm T}$)
with individual time step has been developed.  We incorporate this ${\rm P}^3{\rm T}$ scheme into our code.

In this section we describe the concept and the implementation of our code, \ourcode.

\subsection{Basic equations\label{ss:scheme}}
\subsubsection{The ${\rm P^3T}$ scheme}

The ${\rm P^{3}T}$ scheme \citep{Oshino2011} is a hybrid integrator based on the splitting of the Hamiltonian.  
In this scheme, the Hamiltonian of the system of particles is divided into two parts by the distances between particle pairs. They are called
the soft part and the hard part.
The Hamiltonian used in the ${\rm P^{3}T}$ scheme is given by
\begin{eqnarray}
H&=&H_{\rm Soft}+H_{\rm Hard},\\
H_{\rm Soft}&=&-\sum_{i}^{}\sum_{j>i}^{}\frac{Gm_{i}m_{j}}{r_{ij}}W(r_{ij};r_{\rm out}),\label{soft}\\
H_{\rm Hard}&=&\sum_{i}^{}\left[ \frac{|\bm{p}_{i}|^{2}}{2m_{i}}-\frac{GM_{*}m_{i}}{r_{i}} \right]\nonumber\\
&&-\sum_{i}^{}\sum_{j>i}^{}\frac{Gm_{i}m_{j}}{r_{ij}}\left[ 1-W(r_{ij};r_{\rm out}) \right],\label{hard}\\
\bm{r}_{ij}&=&\bm{r}_{i}-\bm{r}_{j},
\end{eqnarray}
where $G$ is the gravitational constant, $M_*$ is the mass of the central star, and $m_{i} , \bm{p}_{i}$, and $\bm{r}_{i}$ are the mass, the momentum and the position of the $i$th particle, respectively. We did not include the indirect term. 
$W(r_{ij};r_{\rm out})$ is the changeover function for the Hamiltonian.  Though the changeover function is determined by both outer and inner cut-off radii\citep[see][]{Oshino2011}, we can express this function by the outer cut-off radius alone. 
This is because the inner cut-off radius is set as $r_{\rm in}=\gamma r_{\rm out}$, where $r_{\rm in}$ and $r_{\rm out}$ are the inner and outer cut-off radii and $\gamma$ is a constant parameter in the range of 0 to 1.

The forces derived from the Hamiltonian are given by
\begin{eqnarray}
\bm{F}_{{\rm Soft},i}&=&-\frac{\partial H_{\rm Soft}}{\partial \bm{r}_{i}}\nonumber\\
&=&-\sum_{j\not=i}^{}\frac{Gm_{i}m_{j}}{{r_{ij}}^{3}}K(r_{ij};r_{\rm out})\bm{r}_{ij},\label{softforce}\\
\bm{F}_{{\rm Hard},i}&=&-\frac{\partial H_{\rm Hard}}{\partial \bm{r}_{i}}\nonumber\\
&=&-\frac{GM_{*}m_{i}}{{r_{i}}^{3}}\bm{r}_{i}\nonumber\\
&&-\sum_{j\not=i}^{}\frac{Gm_{i}m_{j}}{{r_{ij}}^{3}}\left[ 1-K(r_{ij};r_{\rm out}) \right]\bm{r}_{ij},
\label{hardforce}
\end{eqnarray}
where $K(r_{ij};r_{\rm out})$ is the changeover function for the force, defined by
\begin{eqnarray}
W(r)=r\int_{r}^{\infty}\frac{K(\bar{r})}{\bar{r}^{2}}d\bar{r}.
\end{eqnarray}
The changeover function $K(r_{ij};r_{\rm out})$ is determined so that it becomes zero when $r_{ij} < r_{\rm in}$ and unity when $r_{ij} > r_{\rm out}$.  
In \ourcode, we use the same changeover functions as \texttt{PENTACLE} \citep{Iwasawa2017}, which is defined by
\begin{eqnarray}
W(y;\gamma)=\left\{
\begin{array}{ll}
\frac{7(\gamma^6-9\gamma^5+45\gamma^4-60\gamma^3\ln\gamma-45\gamma^2+9\gamma-1)}{3(\gamma-1)^7}y & (y<\gamma)\\
f(y;\gamma)+\left[1-f(1;\gamma)\right]y&(\gamma\leq y<1)\\
1&(1\leq y)
\end{array}\right. ,\nonumber\\
\label{cutfuncW}
\end{eqnarray}
where
\begin{eqnarray}
f(y;\gamma)&=&\Bigl\{ -10/3y^7 + 14(\gamma+1)y^6 -21(\gamma^2+3\gamma+1)y^5\nonumber\\
&&\hspace{2mm}+\left[ 35(\gamma^3+9\gamma^2+9\gamma+1)/3 \right]y^4-70(\gamma^3+3\gamma^2+\gamma)y^3\nonumber\\
&&\hspace{2mm}+210(\gamma^3+\gamma^2)y^2-140\gamma^3y\ln y\nonumber\\
&&\hspace{2mm}+\left.(\gamma^7-7\gamma^6+21\gamma^5-35\gamma^4)\Bigr\}\right/ (\gamma-1)^7.
\end{eqnarray}

Since the changeover function becomes unity when $r_{ij} > r_{\rm out}$, 
gravitational interactions of the hard part work only between particles within the outer cut-off radius.  We call particles within the outer cut-off radius ``neighbors.''
Hence, to integrate the hard part, it is sufficient to consider clusters composed of neighboring particles, which we call ``neighbor clusters''.
We will explain the procedure of time integration and the definition of neighbors and neighbor clusters in Section \ref{ss:procedure}.

The Hamiltonian equation of motion is written as
\begin{eqnarray}
\frac{{d}w}{{d}t}=\left\{w,H\right\},\label{HamiltonEq}
\end{eqnarray}
where $w$ is the canonical variable in the phase space and $\left\{ , \right\}$ denotes Poisson bracket.
The general solution of equation (\ref{HamiltonEq}) at time $t+\Delta t$ from $t$ is written as
\begin{eqnarray}
w(t+\Delta t)=e^{\Delta t \left\{ ,H \right\}}w(t)
\end{eqnarray}
In the ${\rm P^3T}$ scheme, the general solution is approximated as
\begin{eqnarray}
w(t+\Delta t)=e^{\Delta t/2 \left\{ ,H_{\rm Soft} \right\}}e^{\Delta t \left\{ ,H_{\rm Hard} \right\}}e^{\Delta t/2 \left\{ ,H_{\rm Soft} \right\}}w(t).\label{HamiltonSol}
\end{eqnarray}

\begin{figure*}[htbp]
\centering
\includegraphics[width=0.6\linewidth, bb=0 0 556 501]{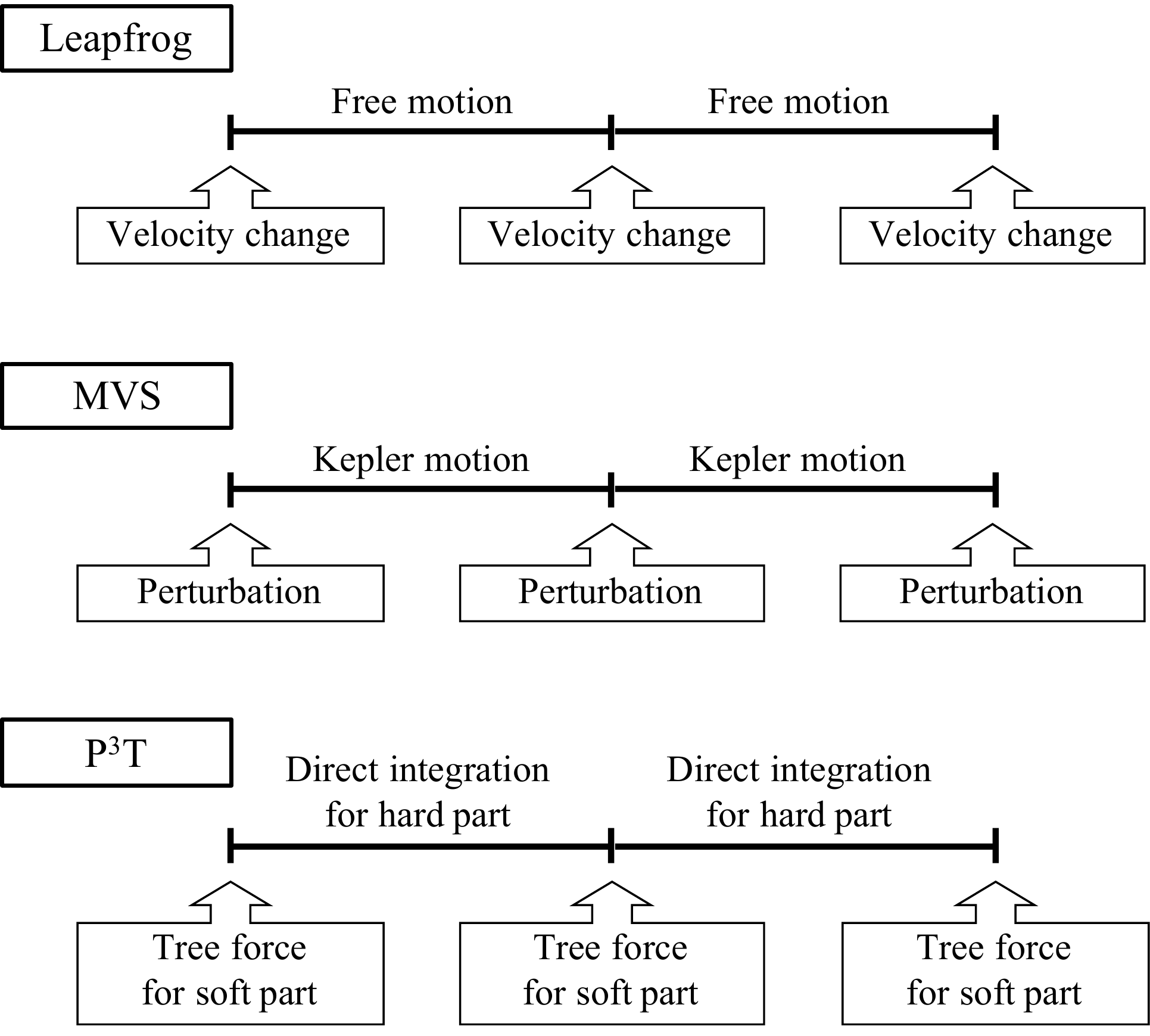}
\caption{
Procedures of thr leapfrog (top), MVS (middle) and ${\rm P^{3}T}$ schemes (bottom). Modified from figure1 in \citet{Fujii2007}.
}
\label{fig:procedureP3T}
\end{figure*}

The ${\rm P^3T}$ scheme adopts the concept of the leapfrog scheme. 
An image of the procedures of the ${\rm P^3T}$ scheme is shown in figure \ref{fig:procedureP3T}. 
In the leapfrog scheme, the Hamiltonian is split into free motion and gravitational interactions. In the mixed variable symplectic (MVS) scheme, it is split into Kepler motion and interactions between particles. In the ${\rm P^3T}$ scheme, it is split into the hard part and the soft part.
The hard part consists of motion due to the central star and short-range interactions. The soft part consists of long-range interactions.
Calculation of the gravitational interactions of the soft part is performed using the Barnes-Hut tree scheme \citep{Barnes1986} available in FDPS \citep{Iwasawa2016}.
Time integration of the hard part is performed using the fourth-order Hermite scheme \citep{Makino1991} with the individual time step scheme \citep{Aarseth1962} for each neighbor cluster or by solving the Kepler equation with Newton--Raphson iteration for particles without neighbors.

Here we explain how we determine the (outer) cut-off radius $r_{\rm out}$.
First, we explain the method used to determine the cut-off radius in previous ${\rm P^3T}$ implementations such as \texttt{PENTACLE}.
Our new method is explained in sub-sections \ref{cutoffind} and \ref{cutoffrandom}.

The cut-off radius is determined based on the Hill radius of each particle, which is defined as  
$r_{{\rm Hill},i}=\left[m_i/(3M_*)\right]^{1/3}a_i$.
Here, $a_i$ is the orbital semi-major axis of the particle.  
The cut-off radius of the $i$th particle is given by
\begin{eqnarray}
r_{{\rm out},i} = \tilde{R}_{{\rm cut},0}r_{{\rm Hill},i},\label{routi}
\end{eqnarray}
where $\tilde{R}_{{\rm cut},0}$ is a parameter.  
If we use a fixed value for all gravitational interactions as the cut-off radius, the cut-off radius used in (\ref{soft}), 
(\ref{hard}), (\ref{softforce}) and (\ref{hardforce}) can be written as
\begin{eqnarray}
r_{\rm out} = \max_{k}\left( r_{{\rm out},k} \right).\label{rout1}
\end{eqnarray}

We call the use of equation (\ref{rout1}) the ``shared cut-off'' method.

\subsubsection{The ${\rm P^3T}$ scheme with individual cut-off\label{cutoffind}}

Particles can have different cut-off radii. 
In our new scheme, 
these different values are actually used for different particles.

The cut-off radius for gravitational interactions between the $i$th and $j$th particles is given by
\begin{eqnarray}
r_{{\rm out},ij} = \max\left( r_{{\rm out},i}, r_{{\rm out},j}\right).\label{rout2}
\end{eqnarray}
We call the use of equation (\ref{rout2}) as the ``individual cut-off'' method. 

The parameter $\tilde{R}_{{\rm cut},0}$ should satisfy $\tilde{R}_{\rm cut} \geq  1$ if we have to ensure that gravitational interactions with particles closer than the distance of the Hill radius are included in the hard part or $\gamma \tilde{R}_{\rm cut} \geq 1$ if we have to ensure that the whole of the gravitational force exerted by particles closer than the Hill radius is calculated in the hard part.  Using the individual cut-off method, 
we have made it possible to split gravitational interactions efficiently,
and to make $\tilde{R}_{{\rm cut},0}$ relatively large 
($\tilde{R}_{{\rm cut},0}\gtrsim 1$) without reducing the simulation speed.
This method requires some complex procedures when two particles with 
different cut-off radii collide and merge.  We will explain the detail 
of this procedure in section \ref{ss:col}.

\subsubsection{The ${\rm P^3T}$ scheme with Hill radius and random velocity dependent cut-off\label{cutoffrandom}}

The cut-off radius should be chosen so that it is sufficiently larger than $v_{\rm ran}\Delta t$, where $v_{\rm ran}$ is the random velocity of particles and $\Delta t$ is the time step for the soft part.
The time step should be sufficiently shorter than the time for the particles to move a distance equivalent to their cut-off radii.

In \ourcode, instead of (\ref{routi}) , the cut-off radius for each particle is set to be
\begin{eqnarray}
r_{{\rm out},i} = \max\left(\tilde{R}_{{\rm cut},0}r_{{\rm Hill},i},
\tilde{R}_{{\rm cut},1}v_{{\rm ran},i}\Delta t\right)\label{routi2}
\end{eqnarray}
where $\tilde{R}_{{\rm cut},0}$ and $\tilde{R}_{{\rm cut},1}$ are the parameters, and
$v_{{\rm ran},i}$ is the mean random velocity for particles around the $i$th particle, where ``random velocity" means the difference between the velocity of the particle and Kepler velocity.
Here we call the method of equation (\ref{routi2}) the ``Hill radius and random velocity dependent cut-off'' method.

The parameter $R_{{\rm cut},1}$ should be determined
so that it satisfies $R_{{\rm cut},1} \geq  1$ in order to let the cut-off radius be sufficiently larger than the product of the random velocity and the time step.
We usually use $R_{{\rm cut},1} = 8$.

To summarize,
in \ourcode, when we use both individual cut-off and Hill radius and random velocity dependent cut-off methods, the cut-off radius for gravitational interactions between the $i$th and $j$th particles is given by
\begin{eqnarray}
r_{{\rm out},ij} = \max\Bigl(
&&\tilde{R}_{{\rm cut},0}r_{{\rm Hill},i},
\tilde{R}_{{\rm cut},1}v_{{\rm ran},i}\Delta t,\nonumber\\
&&\tilde{R}_{{\rm cut},0}r_{{\rm Hill},j},
\tilde{R}_{{\rm cut},1}v_{{\rm ran},j}\Delta t
\Bigr).\label{rout_sum}
\end{eqnarray}

\subsection{Data structure and time-integration procedure\label{ss:implementation}\label{ss:procedure}}

In \ourcode, 
the data structure for particles, which we call for the soft part, is created 
using a function in FDPS. 
The simulation domain is divided into subdomains, each of which is assigned to one MPI process.  Each MPI process stores the data of the particles which belong to its subdomain.  We call this system of particles the soft system.
A particle in the soft system is expressed in a C++ class, which contains as data the index number, mass, position, velocity, acceleration, jerk, time, time step for the hard part and the number of neighbors.
The acceleration of each particle is split into the soft and hard parts, using equations (\ref{softforce}) and (\ref{hardforce}) respectively.
The jerk of each particle is calculated only for the acceleration of the hard part since jerk is not used in the soft part.
Neighbors of the $i$th particle are defined as particles which 
exert hard-part force on the $i$th particle. 
The definition of neighbors and the process of creating a neighbor list are explained in subsection \ref{ss:neighbor}.

For each time step of the soft part, another set of particles is 
created for the time integration of the hard part.
We call this secondary set of particles the hard system. 
The data of the particles are copied from the soft system to the hard system.
A particle in the hard system has second- and third-order time derivatives of the acceleration and the neighbor list, in addition to the data copied from the soft system.
These ``hard particles'' are split into smaller particle clusters, called ``neighbor clusters.''  
Neighbor clusters are created so that 
for any member of one cluster, all its neighbors are also members of that cluster.
The definition of neighbor clusters and the process of creating neighbor clusters are explained in section
\ref{ss:neighbor}.  
Time integration of the hard part can be performed for each neighbor cluster independently since each hard particle interacts only with particles in its neighbor cluster.

The simulation in \ourcode proceeds as follows (see Fig.\ref{fig:procedureP3T}):

\begin{enumerate}
\renewcommand{\theenumi}{\arabic{enumi}}
\renewcommand{\labelenumi}{(\arabic{enumi})}

\renewcommand{\theenumii}{\roman{enumii}}
\renewcommand{\labelenumii}{(\roman{enumii})}

\item
The soft system is created.  The index, mass, position, and velocity of each particle are set from the initial conditions.
\item
Data of the soft system is sent to FDPS.  FDPS calculates the gravitational interactions of the soft part, and returns the  acceleration of the soft part
$a_{\rm Soft}$. 
The neighbor list of each particle is created.\label{pr:FDPS1}

\item
The first velocity kick for the soft part is given, which means that $a_{\rm Soft}\Delta t/2$ is added to the velocity of each particle, 
where $\Delta t$ is the time step of the soft part.  
\label{pr:kick1}

\item
The neighbor clusters are created.  
If there are neighbor clusters of particles stored in multiple MPI processes, the data for particles contained by it are sent to one MPI process (see section \ref{ss:neighbor}).  
The data of particles are copied from the soft system to the hard system.

\item \label{pr:hardpart}
The time integration of the hard system is performed using OpenMP and MPI parallelization.
\begin{enumerate}
\item
The time integration of each neighbor cluster is performed using the fourth-order Hermite scheme.  
If a particle collision takes place, 
the procedure for the collision is carried out (see Section \ref{ss:col}).
\label{pr:hard}
\item
The time integration of each particle without neighbors is performed by solving the Kepler equation.
\end{enumerate}

\item
The data of particles are copied from the hard system to the soft system.
If there are newly born fragments, 
they are added to the soft system.

\item
The data of the soft system is sent to FDPS.  FDPS returns the acceleration of the soft part $a_{\rm Soft}$ in the same way as in step \ref{pr:FDPS1}.  The neighbor list for each particle is created again.
\label{pr:FDPS2}

\item
The second velocity kick for the soft part is given in the same way as in step \ref{pr:kick1}.
\label{pr:kick2}

\item
If collisions of particles take place in this time step, the colliding particles 
are merged and the cut-off radius and the acceleration of the soft part of all particles 
are recalculated.
\label{pr:end}
\item
Go back to step \ref{pr:kick1}.
\end{enumerate}

\subsection{Neighbor cluster creation procedure\label{ss:neighbor}}

A neighbor list is a list of the indices defined for each particle so that the $i$th particle's neighbor list 
contains the particles indices of the $i$th particle's neighbors.
Here, the $i$th particle's neighbors are defined as the particles
which are within the cut-off radius the $i$th particle during the time step of the soft part.  
Numerically, the $i$th particle's neighbors are defined as the particles within the  ``search radius.''

Here we explain how we create the neighbor list of each particle in \ourcode. 
First, particles which are the candidates for neighbors are listed for each particle 
by determining the search radius using as FDPS function,
\begin{eqnarray}
r_{{\rm search},i} = \tilde{R}_{\rm  search,0} r_{{\rm out},i}
+\tilde{R}_{\rm search,1}v_{{\rm ran},i}\Delta t,\label{rsearch}
\end{eqnarray}
where $\tilde{R}_{\rm  search,0}$ and $\tilde{R}_{\rm  search,1}$ are parameters.
$\tilde{R}_{\rm  search,0}$ is set to unity or a value somewhat larger than unity.  
The second term of (\ref{rsearch}) is 
added to the search radius in order 
to include particles which might come into the region within the cut-off radius during the soft step.  In \ourcode, in the case of the individual cut-off method, the search radius is also determined individually for each particle as well as the cut-off radius.
The search radius concerning the interaction between $i$th and $j$th particles is set to the maximum of the search radii of all particles in the case of the shared cut-off method, or the larger of the search radii of the $i$th and $j$th particles in the case of the individual cut-off method.

Second, particles in the $i$th particle's neighbor list which do not satisfy the following condition are excluded from the neighbor list:
\begin{eqnarray}
\tilde{R}_{\rm  search,2}{r_{{\rm out},ij}} < 
\left| \bm{r}_{ij}+\bm{v}_{ij}\Delta t_{\rm min} \right|,
\label{rsearch1}
\end{eqnarray}
where $\bm{r}_{ij}$ and  $\bm{v}_{ij}$ are the position and velocity of the $i$th particle relative to the $j$th particle, and $\Delta t_{\rm min}$ is chosen so that $\left| \bm{r}_{ij}+\bm{v}_{ij}\Delta t_{\rm min} \right|$ takes a maximum $(0<\Delta t_{\rm min} < \Delta t)$.
It can be calculated as:
\begin{eqnarray}
t_{\rm min}=\left\{
\begin{array}{ll}
0 & (-\bm{r}_{ij}\cdot\bm{v}_{ij}/{{v}_{ij}}^2<0) \\
-\bm{r}_{ij}\cdot\bm{v}_{ij}/{v_{ij}}^2 & 
(0\leq -\bm{r}_{ij}\cdot\bm{v}_{ij}/{v_{ij}}^2<\Delta t)\\
\Delta t & (\Delta t\leq -\bm{r}_{ij}\cdot\bm{v}_{ij}/{v_{ij}}^2)
\end{array}
\right. .
\end{eqnarray}
This condition means that if the minimum distance between the $i$th and $j$th particles is sufficiently greater than the cut-off radius, we exclude them from their neighbor lists.
Because the minimum distance can be smaller than the cut-off radius if the relative acceleration is comparable to or greater than the relative velocity, 
the $j$th particles in the $i$th particle's neighbor list which satisfy $v_{ij} < \tilde{R}_{\rm  search,3}a_{ij}\Delta t/2$ are not excluded from the neighbor list even if they satisfy the condition in equation (\ref{rsearch1}), where $a_{ij}$ is the acceleration of the $i$th particle to the $j$-th particle and $\tilde{R}_{\rm  search,3}$ is a parameter larger than unity.

After the neighbor lists of all particles are created, 
the particles are divided into neighbor clusters so that for all particles in a cluster, all of its neighbors are in the same cluster.  

In order to determine a neighbor cluster, 
a ``cluster index'' is set for each particle uing the following steps:
\begin{enumerate}
\renewcommand{\theenumi}{\roman{enumi}}
\renewcommand{\labelenumi}{\roman{enumi}.}
\item
Initially, the cluster index of each particle is set to be the index of the particle (i.e. $i$ for the $i$th particle).\label{pr:nei_start}
\item
The cluster index of a particle is set to the minimum of the cluster indices of all its neighbors and itself.\label{pr:min}
\item
Step \ref{pr:min} is repeated until
the cluster index of all neighbors and itself become equal.
\label{pr:nei_end}
\end{enumerate}
Particles with the same cluster index belong to the same neighbor cluster. 

\begin{figure*}[htbp]
\begin{center}
\includegraphics[width=0.3\linewidth, bb=0 0 456 469]{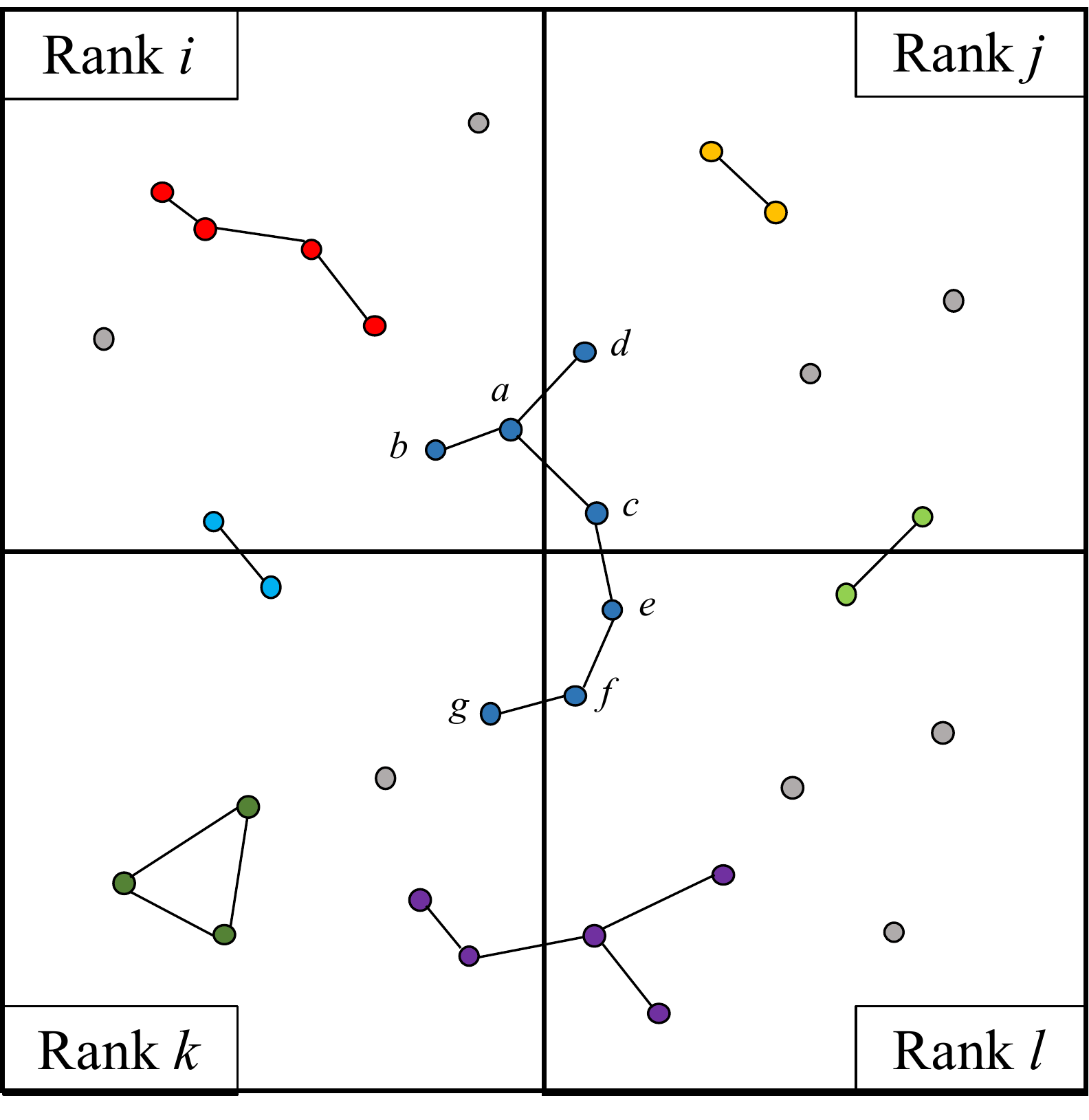}
\hspace{0.02\linewidth}
\includegraphics[width=0.3\linewidth, bb=0 0 457 469]{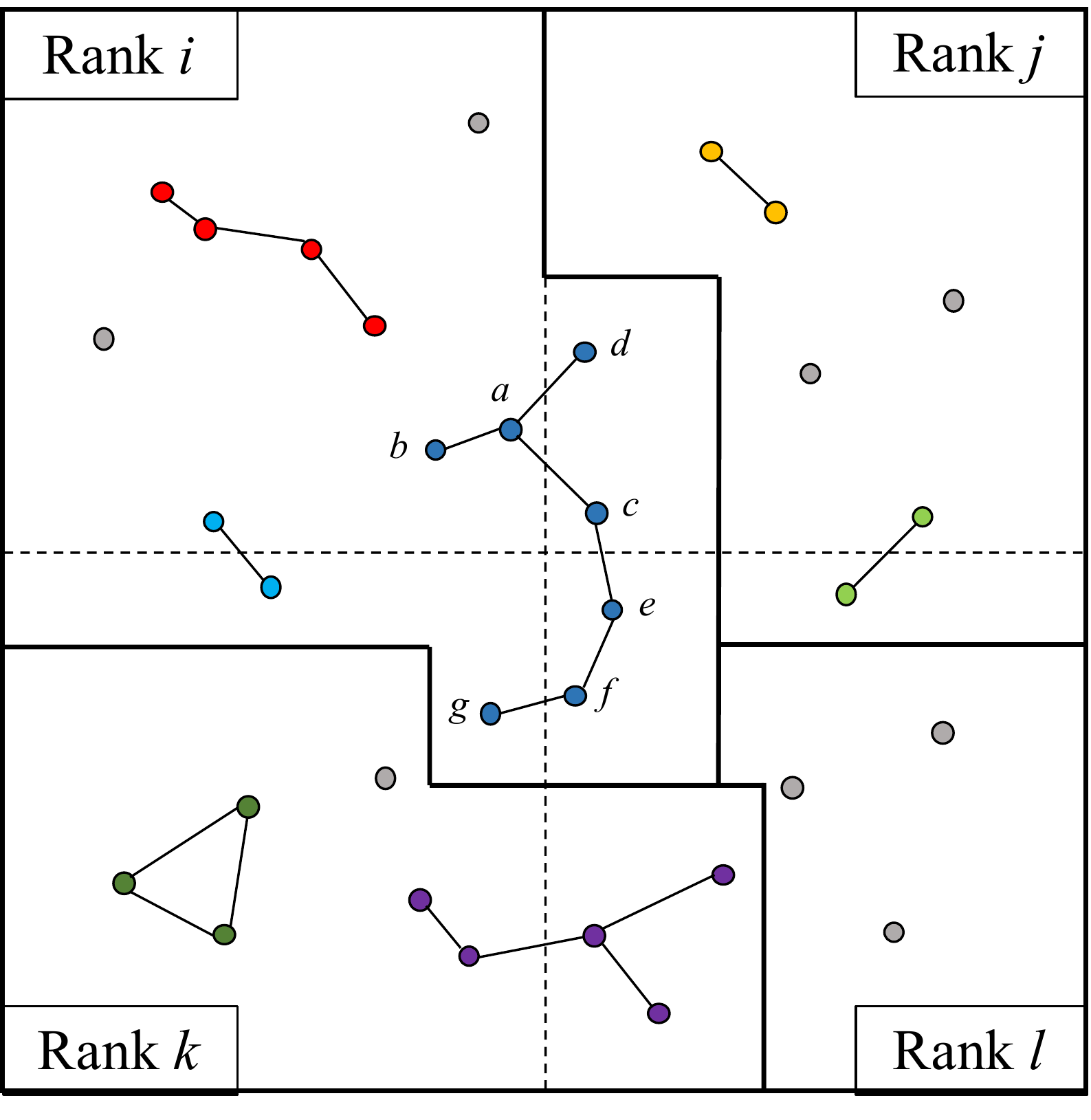}
\end{center}
\caption{
Illustration of particle system and neighbor cluster.
The dots represent particles and the lines connecting particles represent that the connected particles are neighbor pairs.  Particles of the same color belong to the same neighbor cluster (except for the gray particles) and gray particles are isolated particles.  Each divided area represents the area allocated to each MPI process for FDPS (left) and for time integration of the hard part (right).
}
\label{fig:neighbor}
\end{figure*}

If there are neighbor clusters with members from more than one MPI process 
(such as the clusters of blue, cyan, light green and purple particles in figure \ref{fig:neighbor}), the data of particles which belong to such clusters have to be sent to one MPI process 
so that they can be integrated without the need for communication between MPI processes.
Here we explain the procedure to determine the MPI process to which the particle data is sent for each neighbor cluster with members from more than one MPI process.  
Some particles have neighbors from a MPI process different from their own.
Here we call neighbors stored in a different MPI process ``exo-neighbors''.  The set of rank numbers of MPI processes of neighbors including itself
for the $i$th particle is $R_i$.  
For each particle with  exo-neighbors, the procedure to construct the neighbor cluster is as follows:
\begin{enumerate}
\renewcommand{\theenumi}{\Alph{enumi}}
\renewcommand{\labelenumi}{\Alph{enumi}.}
\item
For each particle with exo-neighbors, the cluster index and $R_i$ are exchanged with exo-neighbors of that particle.  Then, the cluster index number is set to be the minimum value of the cluster index numbers of all its exo-neighbors and itself, and $R_i$ is updated to the union of $R_j$ of all its 
exo-neighbors and its $R_i$.  
\label{pr:sum}
\item
Step \ref{pr:sum} is repeated until, the cluster index of all exo-neighbors and itself become equal, and the $R_j$ of all its exo-neighbors and its $R_i$ become equal.\label{pr:nei_fin}
\end{enumerate}
After this procedure, for each nighbor cluster, the time integration will be done on the MPI process which has the minimum rank in $R_i$. Thus, particle data are sent to that MPI process.

\begin{figure*}[htbp]
\begin{center}
\includegraphics[width=0.6\linewidth, bb=0 0 410 215]{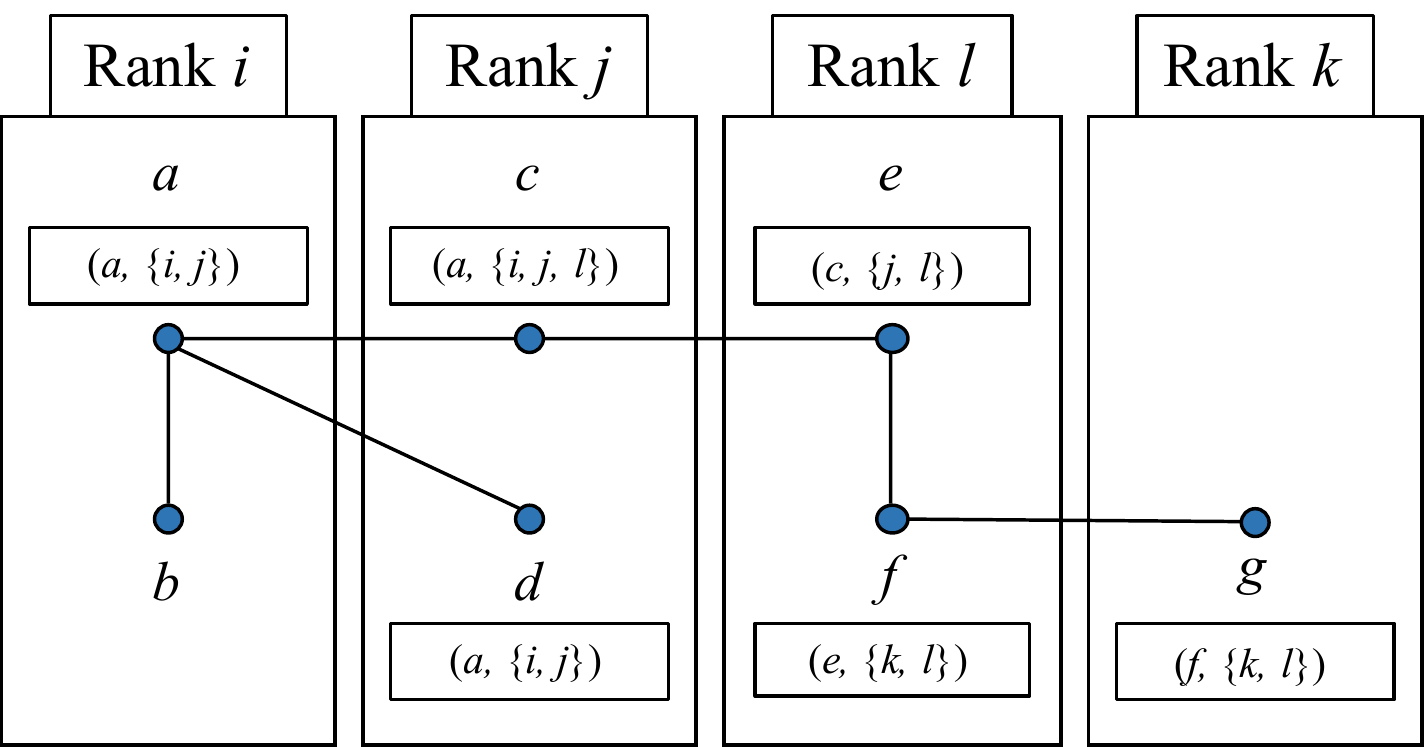}
\end{center}
\caption{
Illustration of neighbor clusters of particles $a$ to $g$ in Fig.\ref{fig:neighbor} before step \ref{pr:sum}.  Here, $a$ to $g$ are the index numbers of each particle assuming that $a<b<c<d<e<f<g$.
The value in the square under the index number indicates $({\rm cluster\ index\ number}, R_i)$ the number the particle has before step \ref{pr:sum}. 
}
\label{fig:exchangeneighbor}
\end{figure*}

Figures \ref{fig:neighbor} and \ref{fig:exchangeneighbor} illustrate these procedures.  In the state shown in the left panel of figure \ref{fig:neighbor}, the neighbor clusters of blue, cyan, light green 
and purple particles span over multiple MPI processes.  In order to integrate the hard part of each neighbor cluster in one MPI process, the allocation of neighbor clusters to each MPI process should be like the right panel of figure \ref{fig:neighbor}.  
Consider the neighbor cluster of particles $a$ to $g$ in figure \ref{fig:neighbor}.
Here we assume that $a<b<c<d<e<f<g$ and $i<j<k<l$.  In this case, after repeating step 
\ref{pr:sum} three times, particles $a$ to $g$ all have $a$ as their cluster index number and $\{i,\ j,\ k,\ l\}$ as $R_i$.
Therefore, data for particles $c$ to $g$ are sent to MPI process $i$ and MPI process $i$ receives data from the rank $j$ to $k$ MPI processes.

Note that our procedure described above is designed to produce no single bottleneck and achieve reasonable load balancing between processes. The communication to construct the neighbor clusters is limited to point-to-point communications between neighboring processes (no global communication), and
the time integration is also distributed to many MPI processes.

\subsection{Treatment of collisions\label{ss:col}}

\subsubsection{Perfect accretion model}

Here we explain the procedure for handling collisions for the case of the perfect accretion model.

The procedure for handling collision is performed during the time integration of the hard part (step \ref{pr:hard}, see subsection \ref{ss:procedure}).  Two particles, which we call the $i$th and $j$th 
particles, are considered to have collided when
\begin{eqnarray}
r_{ij} < f (R_{{\rm p},i}+R_{{\rm p},j}),
\end{eqnarray}
where $R_{{\rm p},i}$ and $R_{{\rm p},j}$ are the radii of the $i$th and $j$th particles respectively.  The coefficient $f$ is the enhancement factor of radius.  If perfect accretion is assumed,  these two particles are replaced by a new particle with mass $m_i+m_j$, where $m_i$ and $m_j$ are the respective masses of the $i$th and $j$th particles.  The position and velocity of the new particle are set so that the position of center of gravity and momentum are conserved:
\begin{eqnarray}
\bm{r}_{\rm new} &=& \frac{m_i \bm{r}_{i}+m_j \bm{r}_{j}}{m_i+m_j},\\
\bm{v}_{\rm new} &=& \frac{m_i \bm{v}_{i}+m_j \bm{v}_{j}}{m_i+m_j}.
\end{eqnarray}

The energy dissipation due to the collision is 
calculated as the summation of the dissipation of the relative kinetic 
energy and gravitational interaction of two particles, and the 
change in the interaction energy with others due to the change in position.
Thus we have
\begin{eqnarray}
E_{\rm disp, Hard}&=&
\varepsilon_0+\varepsilon_1+\varepsilon_2+\varepsilon_3,\label{edisphard}
\end{eqnarray}
\begin{eqnarray}
\varepsilon_0&=& \frac{1}{2}\mu_{ij}|\bm{v}_{ij}|^2, \label{e0}\\
\varepsilon_1&=&- \frac{Gm_im_j}{r_{ij}}W(r_{ij};r_{{\rm out},ij}),\\
\varepsilon_2&=&- {GM_*}\left[\frac{m_i}{r_{i}}+\frac{m_j}{r_{j}}-\frac{m_i+m_j}{r_{\rm new}}\right],\\
\varepsilon_3&=&-Gm_i\sum_{k\in N_i}m_k\left[\frac{W(r_{ik};r_{{\rm out},ik})}{r_{ik}}-\frac{W(r_{{\rm new}k};r_{{\rm out},ik})}{r_{{\rm new}k}}\right]\nonumber\\
&&-Gm_j\sum_{k\in N_j}m_k\left[\frac{W(r_{jk};r_{{\rm out},jk})}{r_{jk}}-\frac{W(r_{{\rm new}k};r_{{\rm out},jk})}{r_{{\rm new}k}}\right],\nonumber\\
\label{e3}
\end{eqnarray}
where $\mu_{ij} = m_im_j/(m_i+m_j)$ is the reduced mass and $N_i$ is the neighbor list of the $i$th particle,
$\varepsilon_0$ represents the dissipation of the relative kinetic energy of two particles, $\varepsilon_1$ the dissipation of gravitational potential between two particles, 
$\varepsilon_2$ the change of gravitational potential with respect to the central star, and 
$\varepsilon_3$ the change of gravitational potential between the neighbors of the $i$th and $j$th particles.
If the changeover functions in (\ref{e0}) to (\ref{e3}) are replaced by unity, 
the sum of $\varepsilon_0$ to $\varepsilon_3$ becomes the energy dissipation of the total of the soft and hard parts.  
Although the gravitational potential of particles other than neighbors also change, they are ignored.  
The accuracy of the simulation can be checked by the error in the total energy, taking into account the dissipations 
mentioned above.

In the case of  individual cut-off, the $i$th and $j$th particles usually have different cut-off radii.  Therefore, the masses $m_i$ and $m_j$ in the new particle are subjected to different hard part forces 
calculated by the different cut-off radii of the $i$th and $j$th particles.
However, the masses $m_i$ and $m_j$ should move together as one particle because they have merged.  
In \ourcode, the new particle (consisting of the $i$th and $j$th particles) is considered as composed of two particles.  
In other words, the $i$th and $j$th particles are not replaced by a new particle during the time integration of the hard part.  
The force on the new particle is calculated in the following steps.
First, 
the hard part accelerations of the $i$th and $j$th particles, $\bm{a}_{{\rm Hard},i}$ and $\bm{a}_{{\rm Hard},j}$, are calculated separately, 
except for the contribution of the interaction between these two particles;
then the hard part acceleration of the new particle is calculated by
\begin{eqnarray}
\bm{a}_{{\rm Hard}} &=& \frac{m_i\bm{a}_{{\rm Hard},i}+m_j\bm{a}_{{\rm Hard},j}}{m_i+m_j}.
\end{eqnarray}
These two particles are replaced by a new particle after the second velocity kick of the soft step (step \ref{pr:end} in subection \ref{ss:procedure});
since the $i$th and $j$th particles have different cut-off radii, they feel different soft forces.
The acceleration for the soft velocity kick is their mass-weighted average.
Thus, there is a small energy dissipation due to this averaging process, expressed as
\begin{eqnarray}
E_{\rm disp, Soft} 
=& \frac{1}{2}\mu_{ij}|\bm{v}_{ij}|^2,
\end{eqnarray}
where $\bm{v}_{ij}$ is the relative velocity of the $i$th and $j$th particles right after the velocity kick is given.  In the soft part, potential energy dissipation is not present since the particle position does not change before and after merging.
After the two particles are merged, the cut-off radius is recalculated.  
The soft and hard part acceleration and jerk of all particles are recalculated since the change of cut-off radius influences both hard and soft parts of the Hamiltonian.

\subsubsection{Implementation of fragmentation\label{ss:fragmentations}}

First, we describe how the fragmentation process is treated in the code.  
The procedure for particle collision with fragmentation is similar to the case of the perfect accretion.  
When a collision occurs, remnant and fragment particles are created.
The number and masses of the remnant and fragments are determined using the fragmentation model.

In \ourcode, as in the case of perfect accretion, mass originating from the $i$th and $j$th particles is considered as separate particles, 
until the end of the hard integration steps.
We assume that the total mass of the fragments is smaller than the mass of the smaller of the two collision participants.
Therefore, we assume that fragments adopt the cut-off radius of the smaller collision participants, and the remnant will be composed of the larger participant and the rest of the mass of the smaller participant.

\subsubsection{Fragmentationl models\label{ss:kom}}

In this section, we describe the fragmentation models  implemented in \ourcode.
Currently, two models are avaiiable. 
One is a very simplified model, which has the advantage that we can study the effect of changing the
collision product. 
The other is a model that can adjust the number of fragments by collision angle and relative velocity based on \citet{Chambers2013}, which determines the collision outcome using the result of smooth-particle hydrodynamic collision experimentation.
In this model, the collision scenario, which includes accretion, fragmentation, and hit-and-run, is also determied by collision angle and relative velocity.
Since the latter is given in \citet{Chambers2013}, in the following we describe the simple model only.

We first present the simple model.
Here, the mass of the remnant is given by $(1-a)m_i+m_j$, 
where $a$ is a parameter in the range of 0 to 1,
and $m_i$ and $m_j$ are the masses of two colliding particles
($m_i \le m_j$).
The mass $am_{i}$ goes to  the fragments.  The number of fragments, $n_{\rm frag}$, is given by
\begin{eqnarray}
n_{\rm frag} &=& \min\left(
\left\lfloor \frac{am_{i}}{m_{\rm min}} \right\rfloor, N_{\rm frag}\right).
\end{eqnarray}
Here, $m_{\rm min}$ is the minimum mass of the particles
and $N_{\rm frag}$ is the maximum number of fragments for one collision.
If $n_{\rm frag}=1$, we set it to 0 and apply the procedure for perfect accretion.
The fragments all have the same mass,
\begin{eqnarray}
m_{\rm frag} &=& \frac{am_{i}}{n_{\rm frag}}.
\end{eqnarray}
The fragments are placed on a circle with center at the position of the remnant on the  plane of the orbital angular momentum of the relative motion of the two particles.  
The velocities of the fragments relative to the remnant are set to be 1.05 times the escape velocity of the remnant.

The energy dissipation in the hard part due to the collision can be calculated in the same manner as for perfect accretion.

\section{Result}

\subsection{Initial conditions and  simulation parameters\label{ss:condition}}

In this section, we present the initial models, parameters and
computing resources and parallelization method used.

For standard runs, we use $10^6$ planetesimals with equal masses of $2 \times 10^{21}\, {\rm g}$ distributed in the region $0.9$--$1.1\, {\rm au}$ from the Sun. 
Therefore, the total mass of solid materials is $2 \times 10^{27}\, {\rm g}$. When we change the total number of particles, the surface mass density is kept unchanged.
The solid mass is consistent with that of the minimum-mass Solar nebula\citep[MMSN;][]{Hayashi1981}.    Initial orbital eccentricities and inclinations of 
planetesimals are given by Gaussian distribution with dispersion 
$\langle e^2 \rangle^{1/2} = 2\langle i^2 \rangle^{1/2}=2h$ 
\citep{Ida1992}, where $h$ is the reduced Hill radius defined 
by $h=r_{\rm Hill}/a$. The Hill radius $r_{\rm Hill}$ is given by
\begin{equation}
r_{\rm Hill} = \left( \frac{m_{\rm p}}{3M_\odot}\right)^{1/3}a.
\end{equation}
The particle density is set to be $2\, {\rm g\, cm^{-3}}$.  
In the wide-range simulations we use $10^6$ planetesimals with equal masses of $1.5 \times 10^{24}\, {\rm g}$ distributed in the region $1.0$--$10\, {\rm au}$ from the Sun. 

We use $\eta = 0.01$ for the  accuracy parameter for the fourth-order
Hermite scheme. For the initial step and also for the  first step
after a collision,
we use $\eta_0 = 0.1\eta$.
We set $\tilde{R}_{\rm search0}=1.1,\,
\tilde{R}_{\rm search1}=6,\, \tilde{R}_{\rm search2}=2$, and
$\tilde{R}_{\rm search3}=2$, (see equations (\ref{rout_sum})).
For the accuracy parameter of the Barnes--Hut algorithm
we use the opening angle of  $\theta = 0.1$ and  0.5.
The system of units is that solar mass, the astronomical unit, and the gravitational constant are all unity. 
In these units, $1\, {\rm yr}$ corresponds to $2\pi$ time units.

The calculations in this paper were carried out on a Cray XC50 system at the Center for Computational Astrophysics (CfCA) of the National Astronomical Observatory of Japan (NAOJ).  
This system consists of 1005 computer nodes, and each node has Intel Xeon Skylake 6148 (40 cores, 2.4\,GHz) processors.  We used MPI over up to 208 processors. 
Unless otherwize noted, OpenMP over five threads and the AVX512 instruction set were used. 
Some of the calculations were done on a Cray XC40 system at the Academic Center for Computing and Media Studies (ACCMS) of Kyoto University.
This system consists of 1800 computer nodes, and each node has Intel Xeon Phi KNL (68 cores, 1.4GHz) processors.
We used MPI over 272 processes, OpenMP of four threads per process,
and the AVX2 instruction set in this system.

We used FDPS version 5.0d.\citep{Iwasawa2016} with the performance enhancement for the
exchange of the local essential tree  \citep{Iwasawaetal2019}.

\subsection{Accuracy and performance \label{s:Performance}}

In sub-subsection \ref{sect:eqmass}, we present the measured accuracy and
performance for the case of equal-mass particles, 
and in section \ref{sect:masspectrum} that for systems with a mass spectrum.
Finally, in sub-subsection \ref{sect:longterm}, we present the result of
long-term calculations.

\subsubsection{Equal-mass systems}
\label{sect:eqmass}

\begin{figure*}[htbp]
\begin{center}
\includegraphics[width=0.8\linewidth, bb=0 0 882 664]{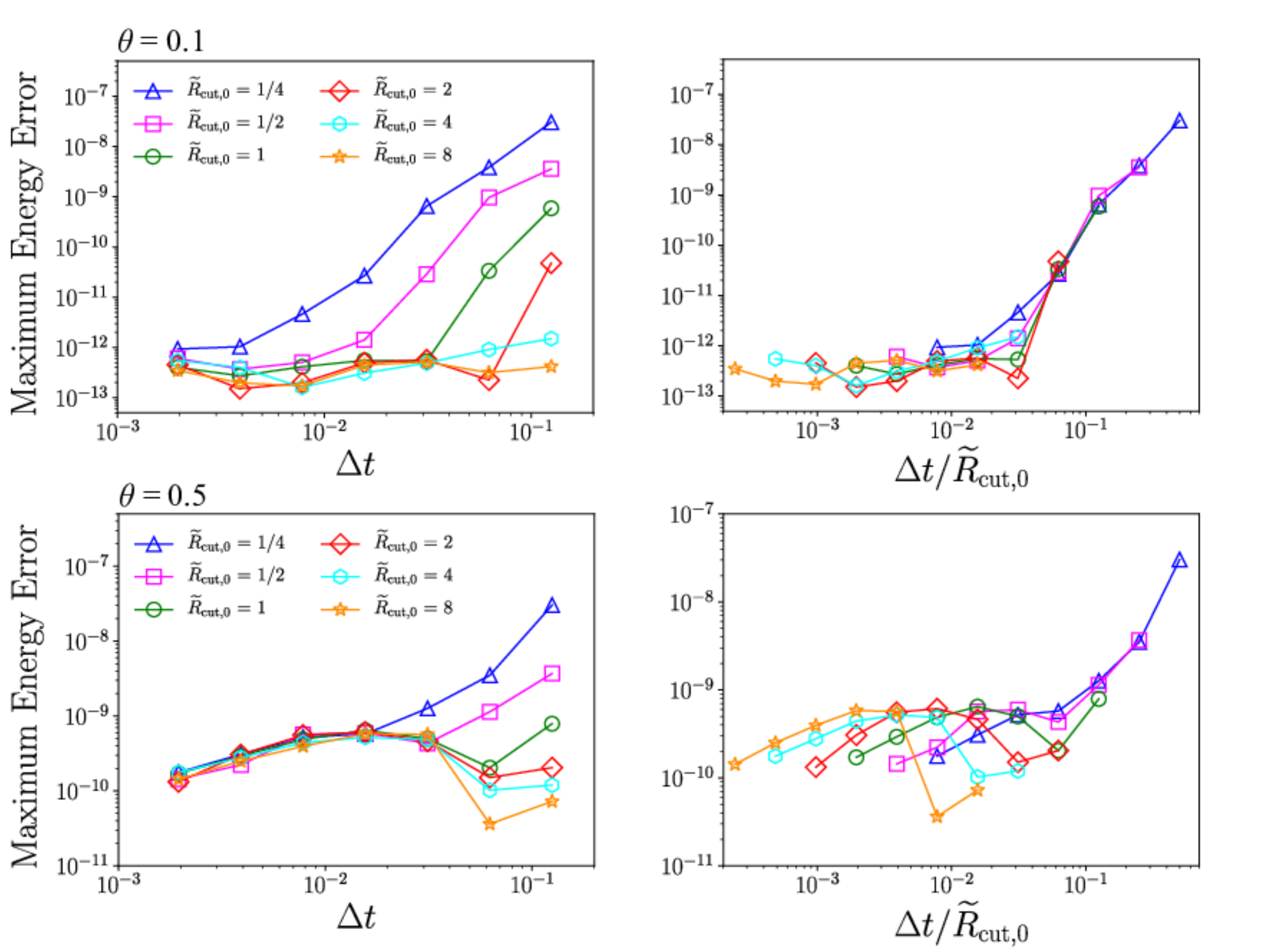}
\end{center}
\caption{
Maximum energy error over 10 Keplerian orbits as functions of $\Delta t$ (left) and $\Delta t/\tilde{R}_{\rm cut}$ (right) in the case of $\theta = 0.1$ (top) and $\theta = 0.5$ (bottom).
}
\label{fig:energy_error_all}
\end{figure*}

In this sub-subsection we present the results of calculations with equal-mass
initial models.  We use the enhancement factor for particle radius of $f=1$.
Figure \ref{fig:energy_error_all} shows the maximum energy error over 10
Keplerian orbits  as a function of $\Delta t$ and $\Delta
t/\tilde{R}_{\rm cut,0}$.
The energy error here is the relative error of the total energy of the
system, with corrections for  dissipations due to accretion and gas
drag when it is included. We have changed the opening angle $\theta$
and the cutoff radius $\tilde{R}_{\rm cut,0}$.
We used individual cut-off in the standard simulation in this section.
In narrow-range simulations, the individual cut-off radii are almost the same as the shared cut-off radius since the particle masses are equal.

For the case of $\theta=0.1$, the energy error is determined by
$\Delta t/\tilde{R}_{\rm cut,0}$, and not by the actual value of $\Delta
t$, as in \citet{Iwasawa2017}. 
The rms value of the random velocity is $2h$. Therefore, $\Delta t$ must be smaller
than $\tilde{R}_{\rm cut,0}$ in order to resolve the changeover
function, and that is the reason why the error is determined by
$\Delta t/\tilde{R}_{\rm cut,0}$. With  $\Delta t/\tilde{R}_{\rm cut,0}<
0.03$, the integration error reaches the round-off limit of $10^{-12}$.

In the case of a larger opening angle, $\theta=0.5$, the limiting error is around $10^{-10}$. This is simply because the acceleration
error of the soft part is larger than that for $\theta=0.1$.
Since the cut-off radius and the change of distance due to the relative motion between particles in one step are approximately proportional to $\tilde{R}_{\rm cut,0}$ and $\Delta t$, respectively, it is considered that the energy error is determined by $\Delta t/\tilde{R}_{\rm cut,0}$.
The reason why the energy error becomes small as $\tilde{R}_{\rm cut,0}$ and $\Delta t$ are large when $\Delta t/\tilde{R}_{\rm cut,0}$ is small could be because the cut-off radius becomes large with $\tilde{R}_{\rm cut,0}$ and $\Delta t$.

\begin{figure*}[htbp]
\begin{center}
\includegraphics[width=0.8\linewidth, bb = 0 0 878 680]{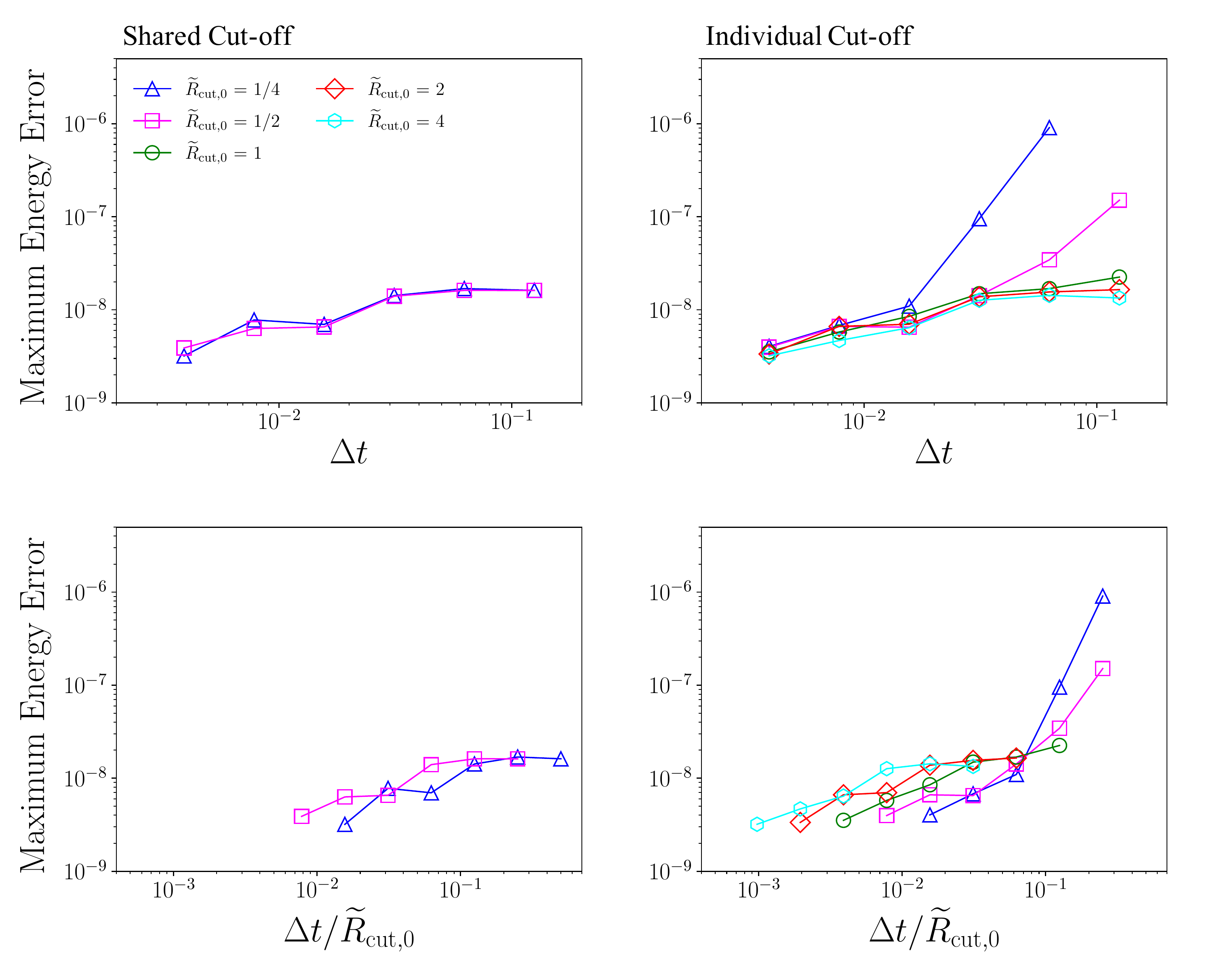}
\end{center}
\caption{
Maximum energy error over 10 Keplerian orbits as functions of $\Delta t$ (left) and $\Delta t/\tilde{R}_{\rm cut}$ (right) in the case of $\theta = 0.5$, shared cut-off (left) and individual cut-off (right) in a wide-range simulation.
}
\label{fig:energy_error_wide}
\end{figure*}

Figure \ref{fig:energy_error_wide} shows the maximum energy error over 10
Keplerian orbits as a function of $\Delta t$ and $\Delta
t/\tilde{R}_{\rm cut,0}$ in a wide-range simulation.
The Keperian orbit in this simulation means that of inner edge.
Only the points where the calculation was completed within 30\, min are plotted.
It shows that the energy error in the case of individual cut-off in a wide-range simulation is not too different from that in the case of shared cut-off if we use $\Delta t/\tilde{R}_{\rm cut,0}\lesssim
0.1$.

\begin{figure*}[htbp]
\begin{center}
\includegraphics[width=0.8\linewidth, bb=0 0 896 347]{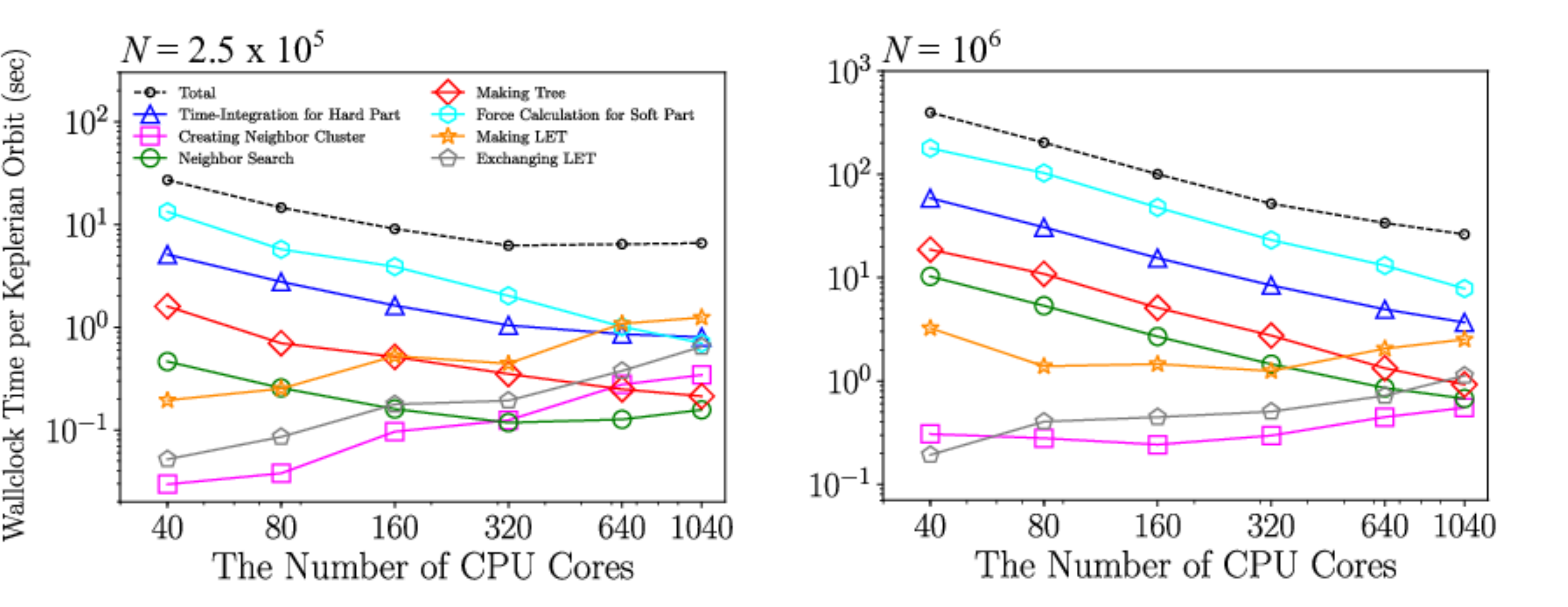}
\end{center}
\caption{
Wallclock time taken for each procedure per Keplerian orbit as a function of the number of CPU cores
in the cases of $1.25\times 10^5$(left) and  $10^6$(right) particles.
We used $\theta =0.5$, $\Delta t = 1/64$ and $\tilde{R}_{\rm cut}=2$.
We used five-thread parallelization(i.e. the number of MPI processes is 1/5 of the number of CPU cores).
}
\label{fig:time_all}
\end{figure*}

Figure \ref{fig:time_all} shows the wallclock time for the integration over one  Kepler time and its breakdown as a function of the number of  CPU cores for the case of $\theta=0.5$,  
$\Delta t=1/64$ and$\tilde{R}_{\rm cut,0}=2$.
We used five cores per MPI process. The wallclock time is the average over ten Kepler times. We can see that the parallel performance speed-up is reasonable for up to 320 cores ($N=1.25\times 10^5$) and more than 1040 cores ($N= 10^6$).

We can see that the times for the soft force calculation, hard part integration, and tree construction all decrease as we increase the number of cores, 
for both $N=1.25\times 10^5$ and $10^6$. 
On the other hand, the times for LET construction, LET communication (exchanging LET),
and creation of the neighbor clusters increase as we increase the number of cores, and the time for LET construction currently limits the parallel speedup. 
LET means local essential tree in FDPS, defined by \citet{Iwasawa2016}.
This increase in the cost of LET construction
occurs because the domain decomposition scheme used in FDPS can result
in suboptimal domains for the case of rings; a simple solution for
this problem is to use cylindrical coordinates\citep{Iwasawaetal2019}
when the ring is relatively nallow. On the other hand, When the radial
range is very wide, the simple strategy used
in \citet{Iwasawaetal2019} cannot be used. We will need some better
solution for this problem.

\begin{figure*}[htbp]
\begin{center}
\includegraphics[width=0.8\linewidth, bb=0 0 896 347]{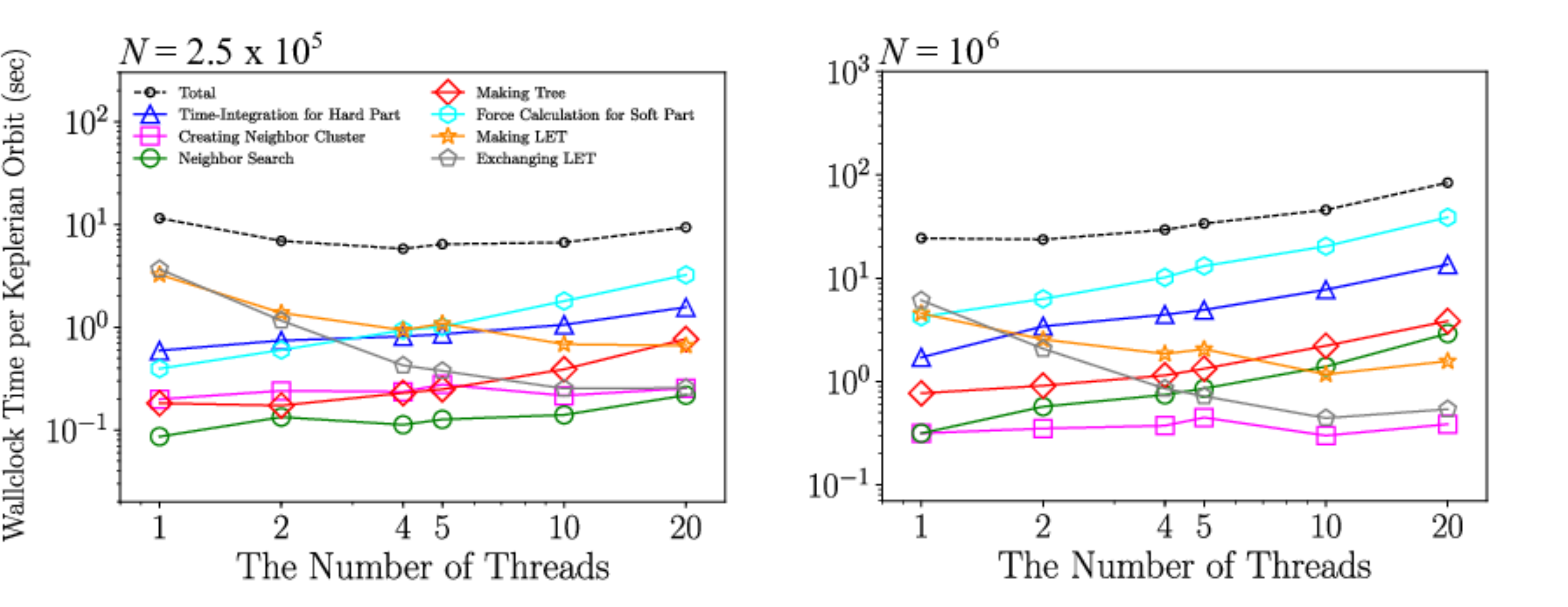}
\end{center}
\caption{
Wallclock time taken for each procedure per Keplerian orbit as a function of the number of MPI processes per node
in the cases of $1.25\times 10^5$(left) and  $10^6$(right) particles.
We used $\theta =0.5$, $\Delta t = 1/64$, $\tilde{R}_{\rm cut}=2$ and 640 CPU cores(i.e. the number of MPI processes is 640 divided by the number of threads).
}
\label{fig:time_thr}
\end{figure*}

Figure \ref{fig:time_thr} shows the wallclock time for 640 cores, but
with different numbers of threads per MPI process. The other parameters are
the same as in figure \ref{fig:time_all}. We can see that the total
time is a minimum for at four threads per process for the case of
$N=1.25\times 10^5$, but at one thread per process for $N=10^6$. This
difference again comes from the costs of the construction and
communication of LETs. With the current domain decomposition scheme,
these costs contain the terms proportional to the number of MPI
processes, and thus for small $N$ and large numbers of MPI processes
these costs can dominate the total cost. Thus, for small $N$,
a combination of OpenMP and MPI tends to give better performance
compared to flat MPI.

\subsubsection{Systems with mass spectrum\label{sect:masspectrum}}

In this sub-subsection we present the results of calculations with particles
with a mass spectrum, in order to evaluate the behavior of \ourcode
at the late stage of planetary formation. As the initial model we used the output snapshot at 9,998 years of integration from
the initial model described in the previous section. The
minimum, average and maximum masses are $2.00\times 10^{21},\, 5.30\times 10^{21}$, and $8.46\times 10^{24}\, {\rm g}$, respectively.
The number of particles is 377740.  We used the enhancement factor for
particle radius of $f=1$.

\begin{figure*}[htbp]
\begin{center}
\includegraphics[width=1\linewidth, bb=0 0 1306 659]{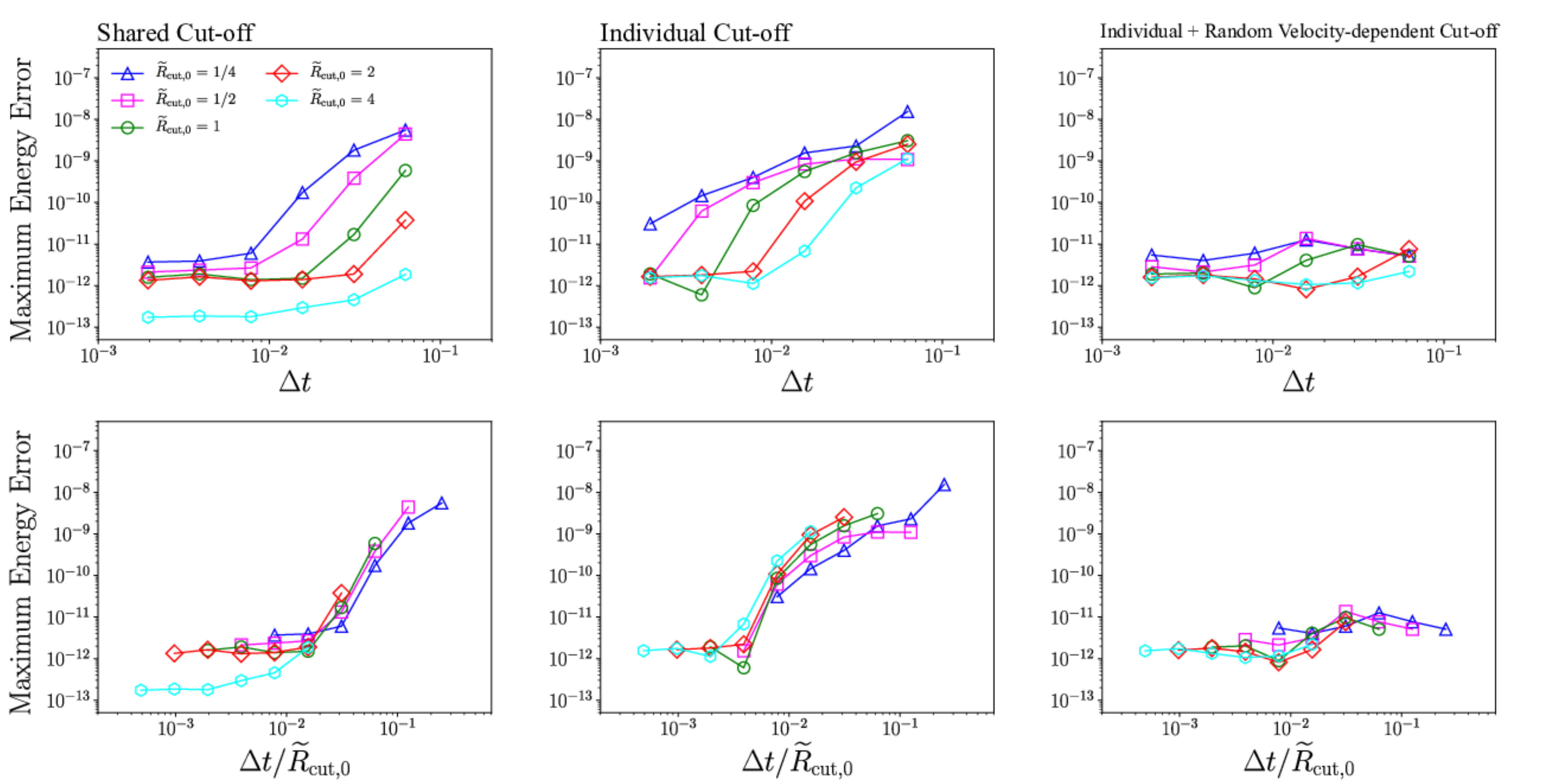}
\end{center}
\caption{
Maximum energy error over five Keplerian orbits as a function of $\Delta t/\tilde{R}_{\rm cut,0}$(top) and $\Delta t$ (bottom) in the case of $\theta = 0.1$, shared cut-off (left), individual cut-off (center), and individual and random velocity-dependent cut-off(right).
}
\label{fig:energy_error_shaind}
\end{figure*}

Figure \ref{fig:energy_error_shaind} shows the maximum energy error over
10 Kepler time as a function of
$\Delta t/\tilde{R}_{\rm cut,0}$ and $\Delta t$ in the case of shared, individual,
and individual and random velocity cut-off schemes. Here, $\theta=0.1$.
If we compare the values of $\Delta t/\tilde{R}_{\rm cut,0}$ itself,
it seems the individual cut-off scheme requires a rather small value of
$\Delta t$, but when the term for the random velocity is included,
we can see that the energy error is essentially independent of $\tilde{R}_{\rm cut,0}$.
Since the energy error which depends on $\Delta t/\tilde{R}_{\rm cut,0}$ when $\Delta t/\tilde{R}_{\rm cut,0}\gtrsim 10^{-2}$ in the shared and individual cut-off schemes does not appear when the random velocity cut-off scheme is used,
this error seems to cause the cut-off radius to not be set sufficiently larger than $v_{\rm ran}\Delta t$. The energy error almost does not depend on $\Delta t$ and $\tilde{R}_{\rm cut,0}$ in the case of the random velocity cut-off scheme.
Therefore, we can use larger $\Delta t$ and smaller $\tilde{R}_{\rm cut,0}$ to reduce the time of simulation while maintaining accuracy.

\begin{figure*}[htbp]
\begin{center}
\includegraphics[width=1\linewidth, bb=0 0 1331 347]{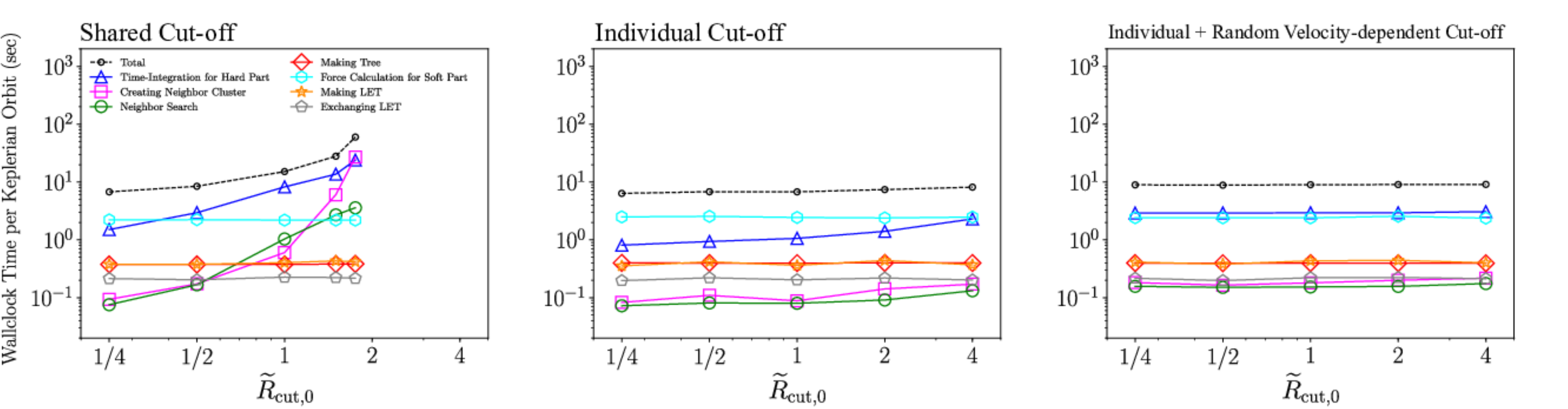}
\end{center}
\caption{
Wallclock time taken for simulations per Keplerian orbit as functions of $\tilde{R}_{\rm cut}$ in the case of $\theta = 0.5$, $\Delta t=1/64$, shared cut-off (left), individual cut-off (center), and individual and random velocity-dependent cut-off(right).
}
\label{fig:time_shaind}
\end{figure*}

Figure \ref{fig:time_shaind} shows the wallclock time for the integration
over one Kepler time and its breakdown as a function of $\tilde{R}_{\rm
cut,0}$ in the case of $\theta =0.5$, $\Delta t = 1/64$.
In the case of the shared cut-off scheme, the calculation cost
increases quickly as we increase $\tilde{R}_{\rm cut,0}$. On the other
hand, from Fig.\ref{fig:energy_error_shaind} we can see that for 
$\Delta t = 1/64$, we need  $\tilde{R}_{\rm cut,0} \ge 1$ in the case
of the shared cut-off scheme to achieve reasonable accuracy.

For individual cut-off schemes with and without the random velocity
term, the total calculation cost is almost independent of
$\tilde{R}_{\rm cut,0}$, and in the case of the scheme with
the random velocity term, the total energy error is also well
conserved for all values of $\tilde{R}_{\rm cut,0}$. Thus, we can see
that individual cut-off schemes with a random velocity term are more
efficient compared to the shared cut-off schemes for realistic
distribution of particle mass and random velocity.

\begin{figure*}[htbp]
\begin{center}
\includegraphics[width=0.45\linewidth, bb=0 0 461 346]{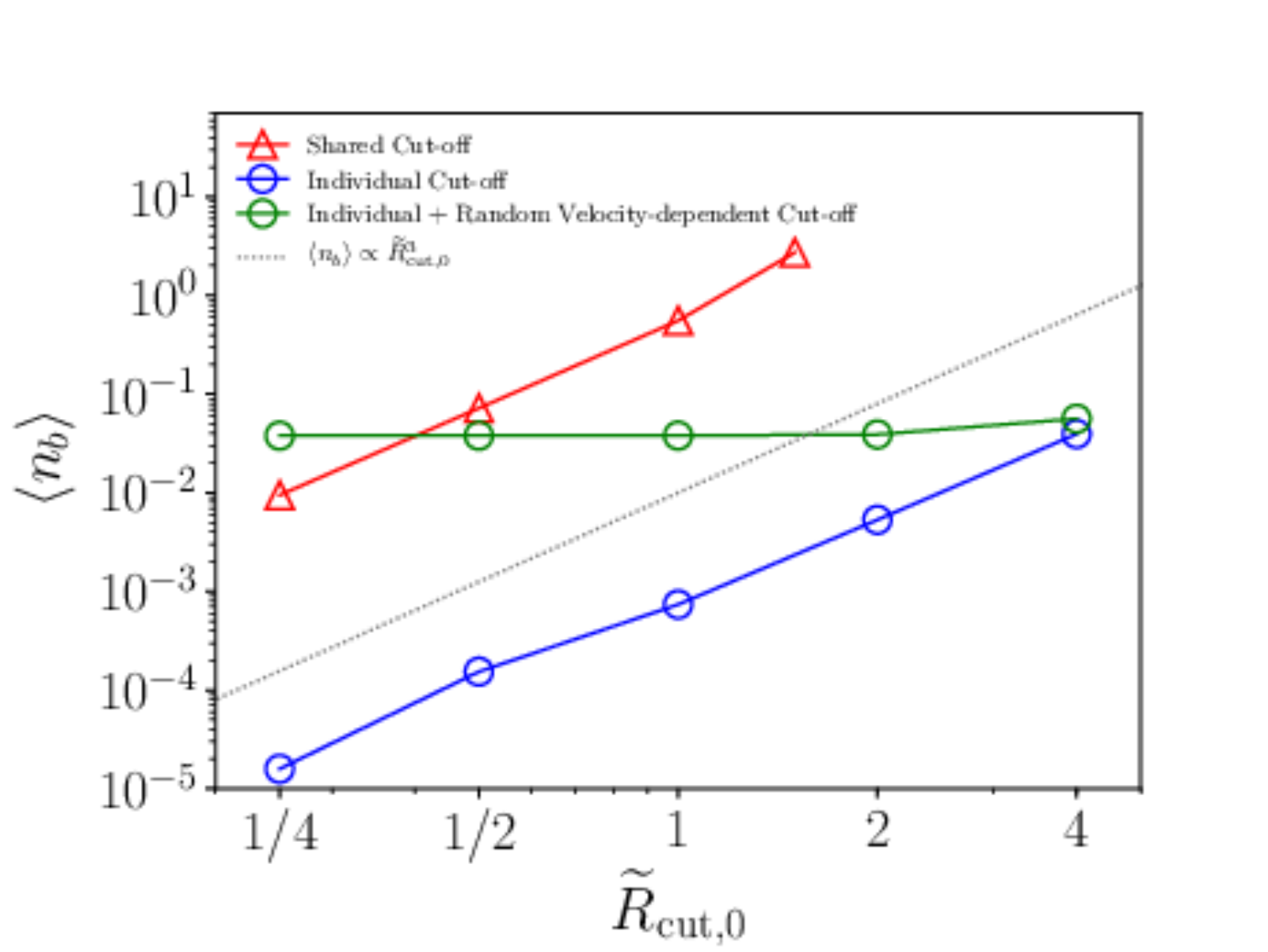}
\includegraphics[width=0.45\linewidth, bb=0 0 461 346]{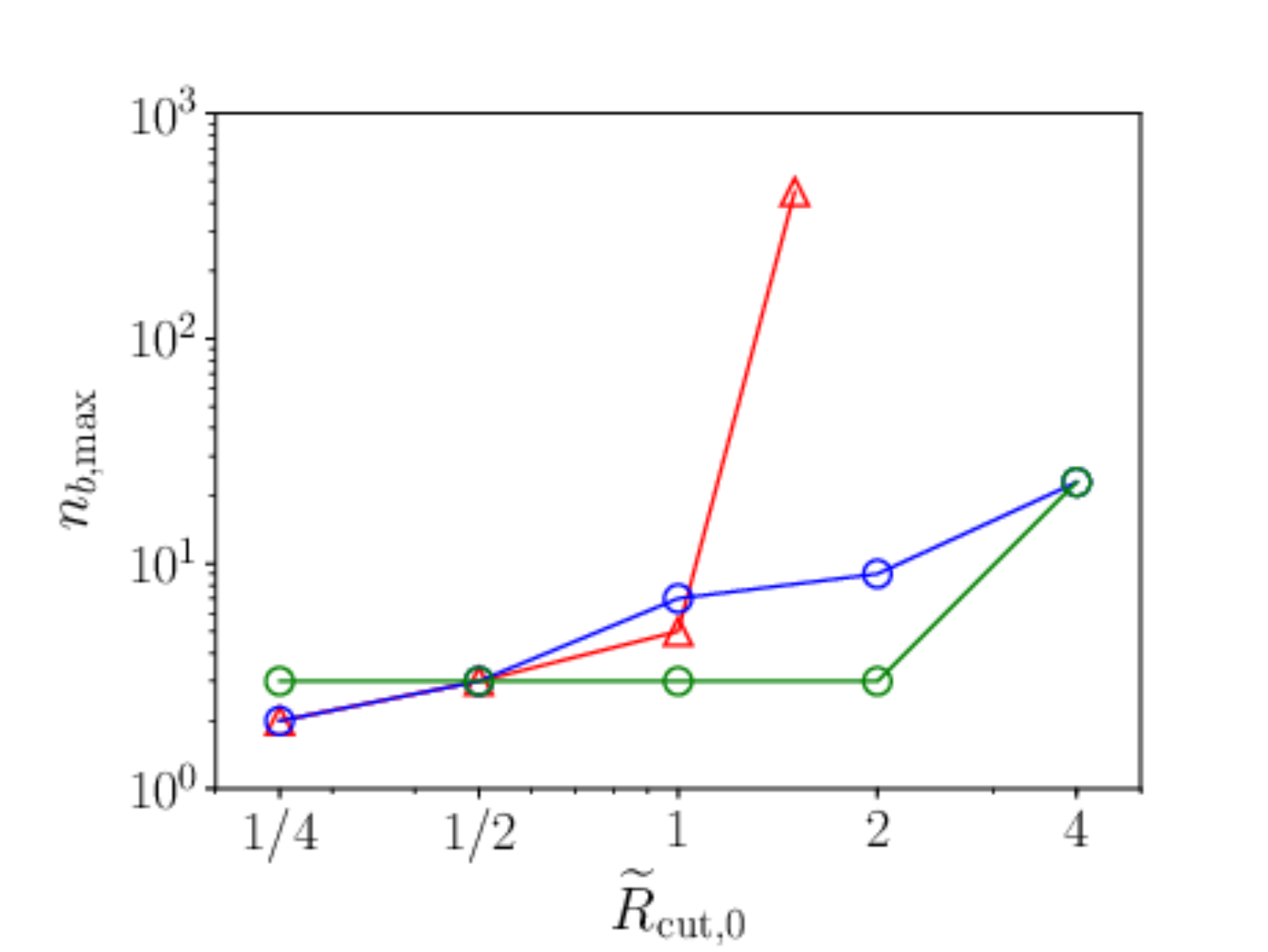}
\end{center}
\caption{
Average number of neighbors for each particle, $\langle n_b\rangle$, (left) and the number of particles in the largest neighbor cluster, $n_{b,{\rm max}}$, (right) as functions of $\tilde{R}_{\rm cut}$ in the case of $\theta = 0.5$, $\Delta t = 1/64$.
The dotted line in the left panel represents the slope of $\langle n_b\rangle\propto\tilde{R}_{\rm cut,0}^3$.
}
\label{fig:cluster_shaind}
\end{figure*}

Figure \ref{fig:cluster_shaind} shows the average number of neighbors,
$\langle n_b\rangle$, and the number of particles in the largest neighbor 
cluster, $n_{b, {\rm max}}$, as functions of $\tilde{R}_{\rm cut,0}$ in the case of $\theta
= 0.5$, $\Delta t = 1/64$. The average number of neighbors is roughly
proportional to ${\tilde{R}_{\rm cut,0}}^3$, while almost independent of ${\tilde{R}_{\rm cut,0}}^3$ for the
case of the individual cut-off with random velocity term. This, of
course, means that for most particles their neighbor is determined
by the random velocity term, and only the neighbors of the most massive
particles are affected by the individual term. This effect on the
neighbors of the most massive particles is very important in maintaining high
accuracy and high efficiency.

\subsubsection{Long-term simulations}
\label{sect:longterm}

In this sub-subsection we present the result of long-term integration of up
to $20000\, {\rm yr}$. We included the gas drag according to the model
in \citet{Adachi1976} for the MMSN model, and we used the simple
fragmentation model with $a=0.3$, $b=0.1$ in the case of the
individual cut-off with random velocity term.
We used parameters of $\theta = 0.5$, $\Delta t = 1/64$,
$\tilde{R}_{\rm cut,0}=3$ and $f=3$. 
We had to stop the simulation with the shared time step since
it had become too slow.

\begin{figure}[htbp]
\begin{center}
\includegraphics[width=1\linewidth, bb=0 0 846 594]{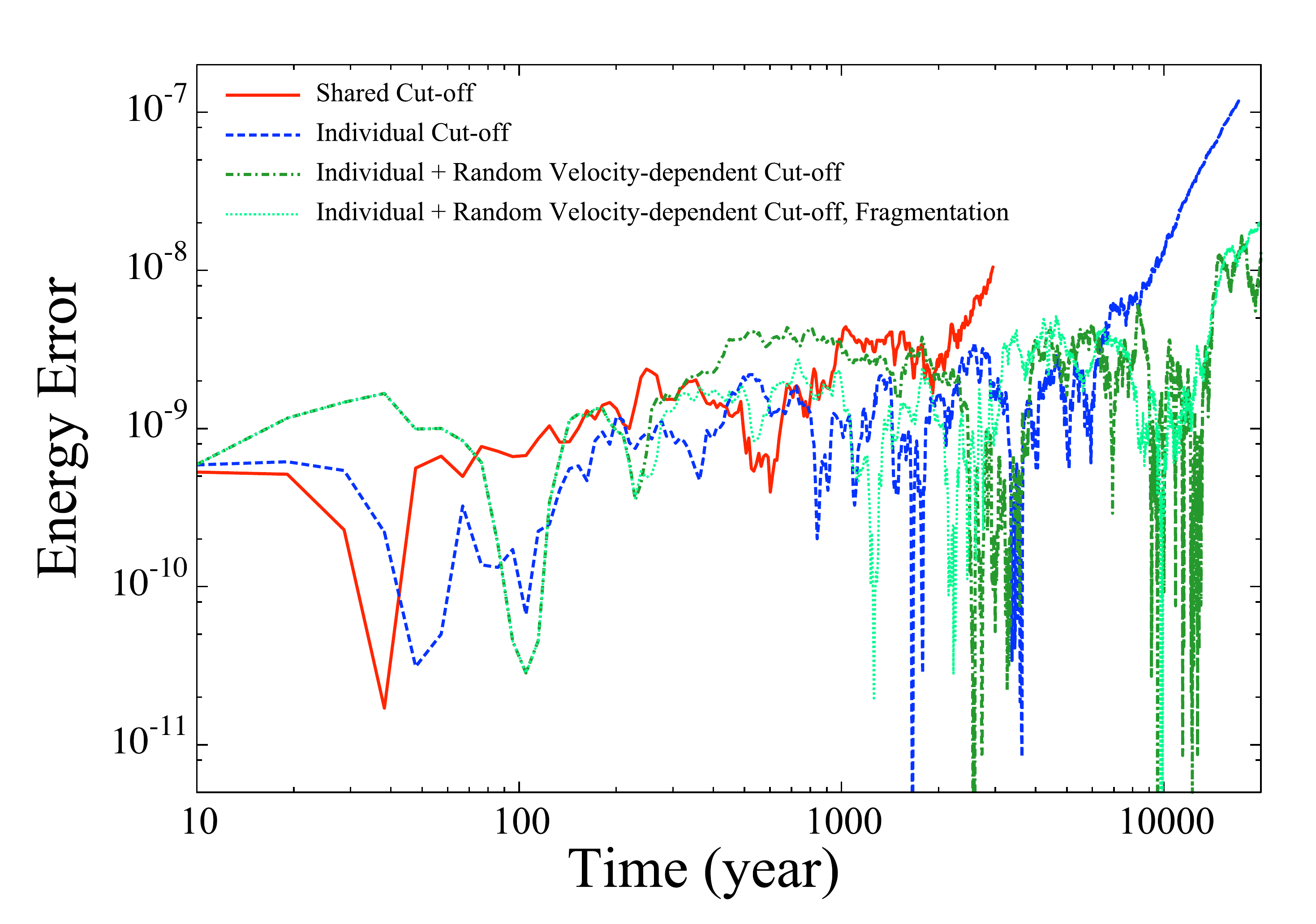}
\end{center}
\caption{
Evolution of energy error for long-time simulations in the case of $\theta = 0.5$, $\Delta t = 1/64$, shared cut-off with perfect accretion (solid red), individual cut-off with perfect accretion (dashed blue),
individual and random velocity-dependent cut-off with perfect accretion (dashed-dotted green),
and  individual and random velocity-dependent cut-off with fragmentation (dotted light green).
}
\label{fig:long_energy_error}
\end{figure}

Figure \ref{fig:long_energy_error} shows the energy error as a function
of time. We can see that for the first $1000\, {\rm yr}$ all the schemes show
similar behavior. However, the error of the run with the shared
cut-off scheme starts to grow by $2000\, {\rm yr}$, and then the calculation
becomes too slow. The error of the run with  the individual cut-off
without the random velocity term also starts to grow by $6000\,{\rm yr}$.
When the random velocity term is included, the error remains small
even after $20000\,{\rm yr}$.
In the case of shared cut-off, it is considered that the energy error due to random velocity appears earlier since the cut-off radius is larger.

\begin{figure*}[htbp]
\begin{center}
\includegraphics[width=0.45\linewidth, bb=0 0 846 594]{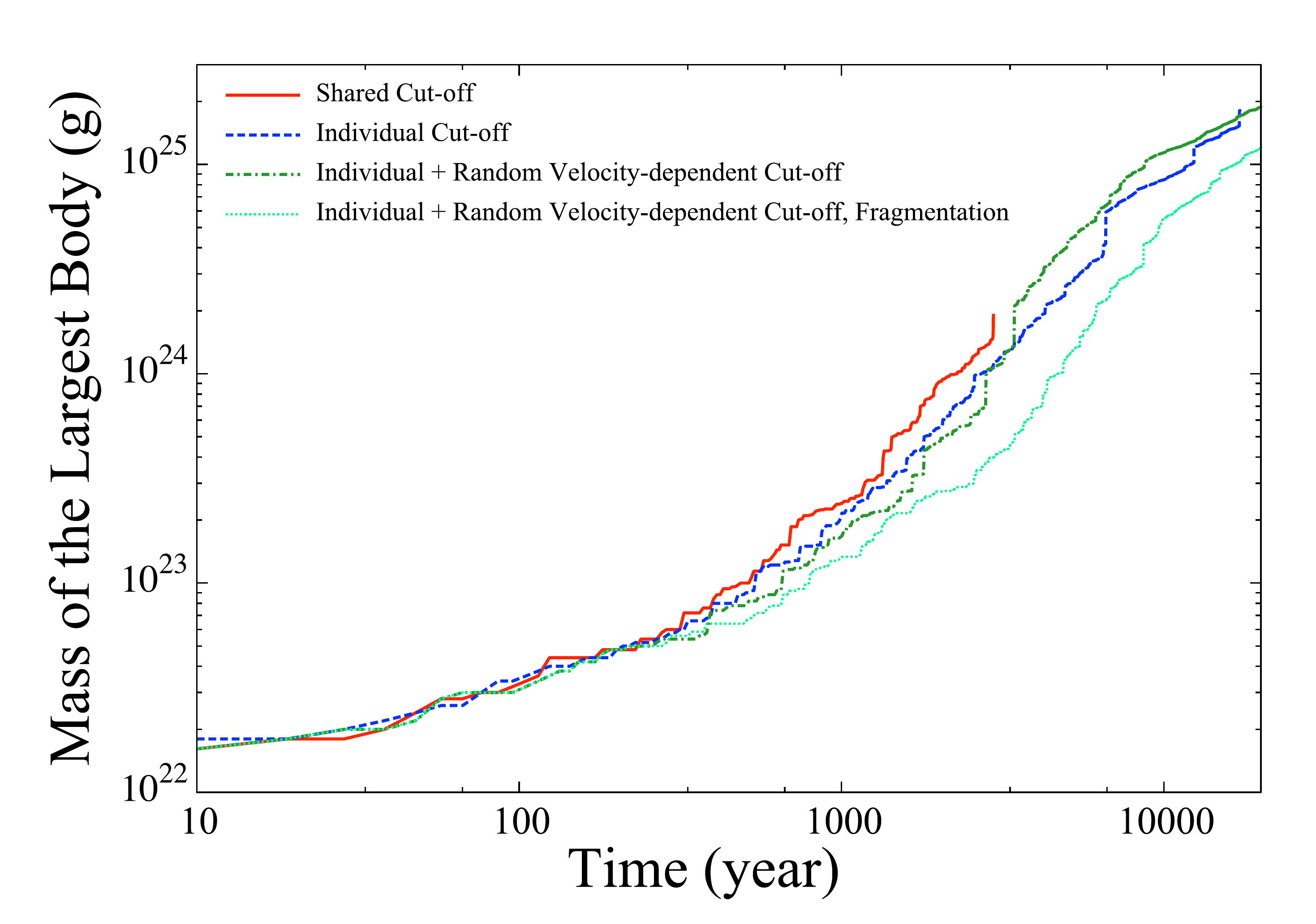}
\hspace{0.02\linewidth}
\includegraphics[width=0.45\linewidth, bb=0 0 846 594]{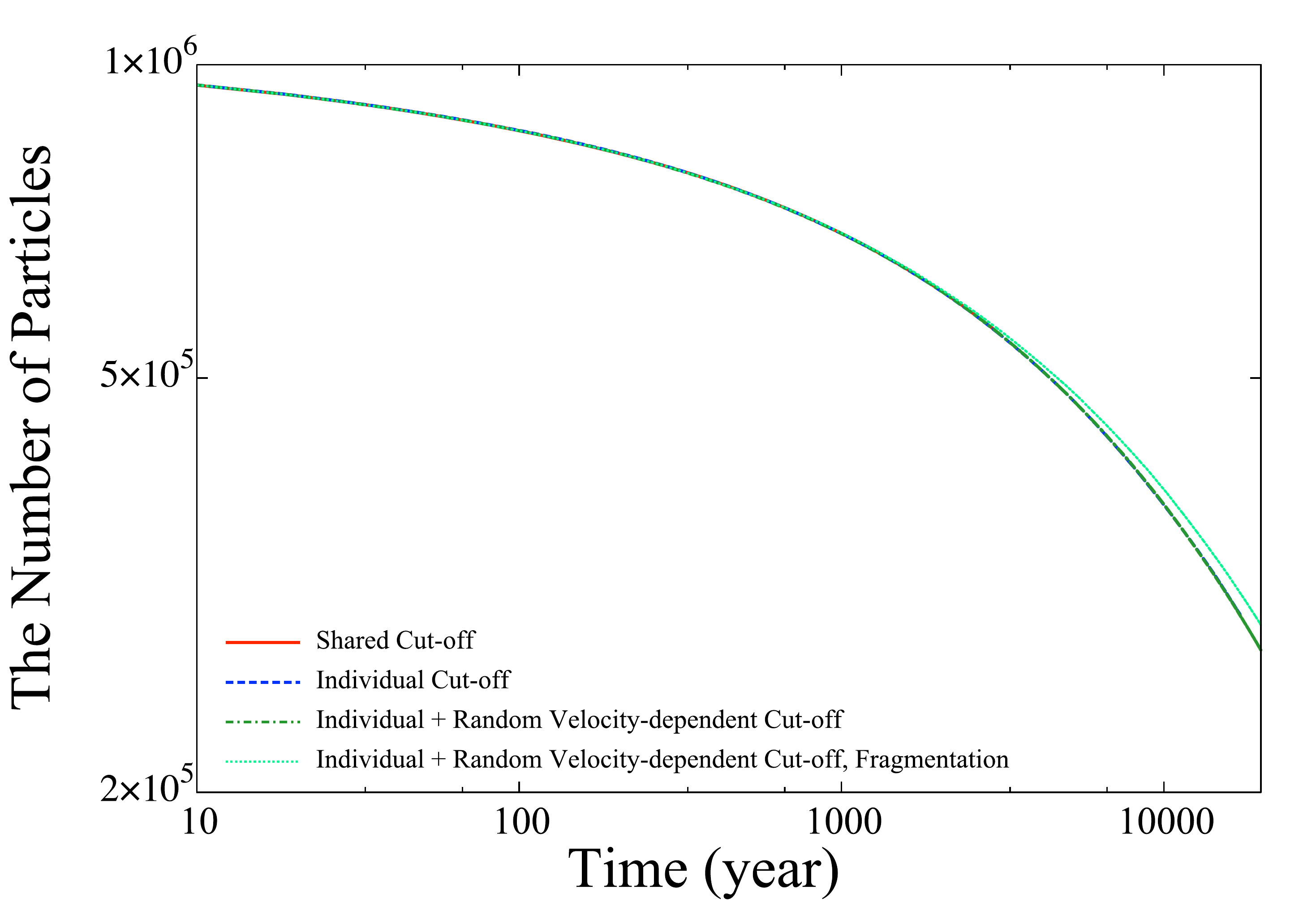}
\end{center}
\caption{
Evolution of number of particles (left) and mass of the largest body (right) for long-term simulations in the case of $\theta = 0.5$, $\Delta t = 1/64$, shared cut-off with perfect accretion (solid red), individual cut-off with perfect accretion (dashed blue),
individual and random velocity-dependent cut-off with perfect accretion (dashed-dotted green),
and  individual and random velocity-dependent cut-off with fragmentation (dotted light green).
}
\label{fig:long_}
\end{figure*}

Note that this result is for one particular choice of the accuracy
parameters and it is possible to improve the error of, for example,
the shared cut-off scheme by reducing the soft time step. On the other
hand,  the individual cut-off scheme with the random velocity term can
keep the error small even after the most massive particle grows by
three orders of magnitude in mass (see figure \ref{fig:long_}).
Thus, we conclude that the the individual cut-off scheme with the
random velocity term can be reliably used for long-term simulations.

\begin{figure}[htbp]
\begin{center}
\includegraphics[width=1\linewidth, bb=0 0 846 594]{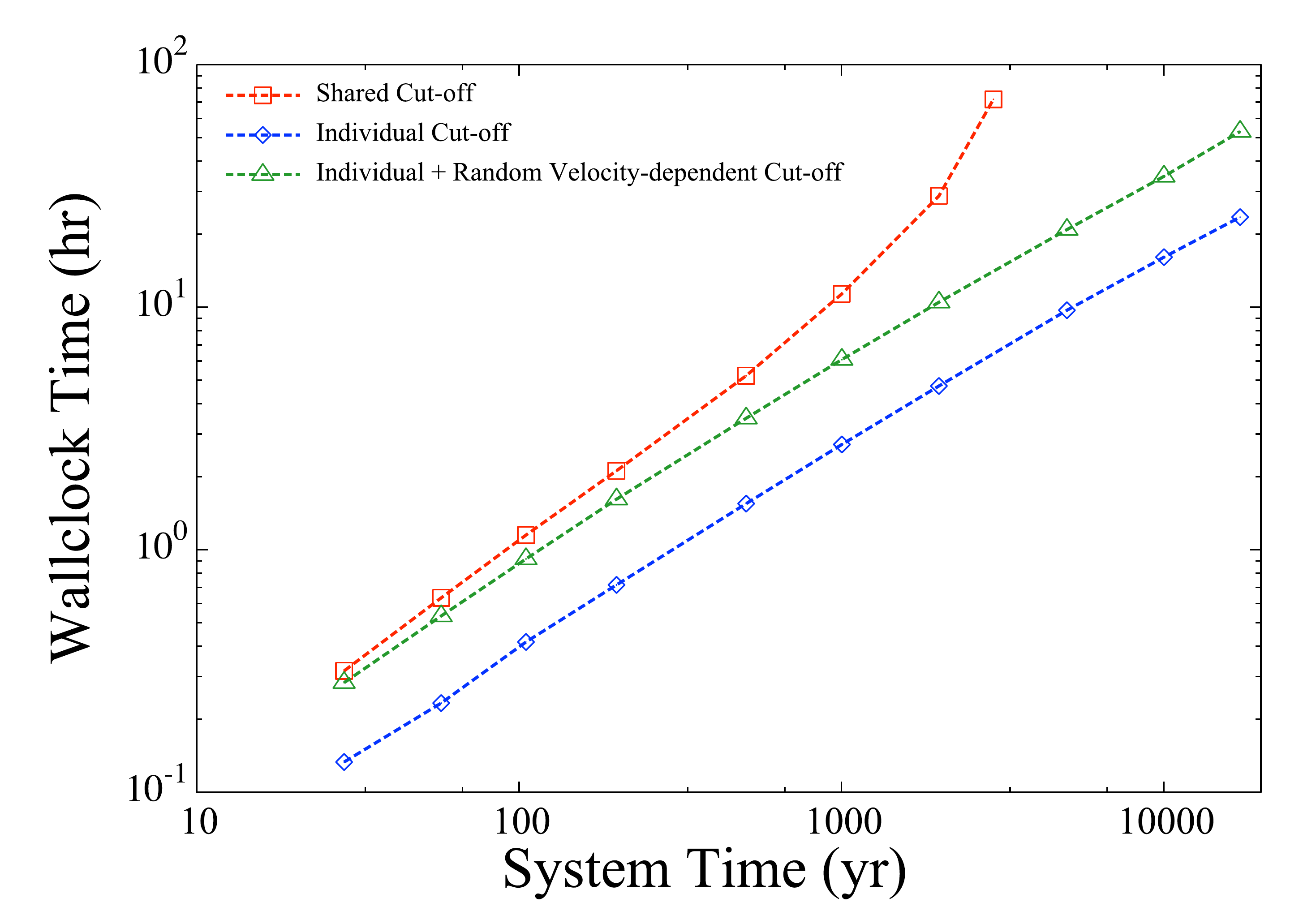}
\end{center}
\caption{
Wallclock time taken for long-term simulations until $t$ as a function of time $t$ in the case of shared cut-off (red) and individual cut-off (blue).
}
\label{fig:wtime}
\end{figure}

Figure \ref{fig:wtime} shows the wallclock time as a function of
simulation time.
The increase of the calculation time of the shared cut-off scheme
is faster than linear, while that of the individual cut-off schemes is
slower, because of the decrease in the number of particles.
At the time of the first snapshot($10\, {\rm yr}$), because the mass of the largest body already reaches about nine times the initial mass, the mean cut-off radius in the case of shared cut-off is about twice as large as for individual cut-off.
This is the reason why the calculation speed in the case of shared cut-off is slower than the individual case from the beginning of the simulation.

Figure \ref{fig:long_} shows the evolution of the number of particles and
the mass of the most massive particle. We can see that the time
evolutions obtained using different cut-off schemes are practically
identical.

\begin{figure*}[htbp]
\begin{center}
\includegraphics[width=0.9\linewidth, bb=0 0 867 347]{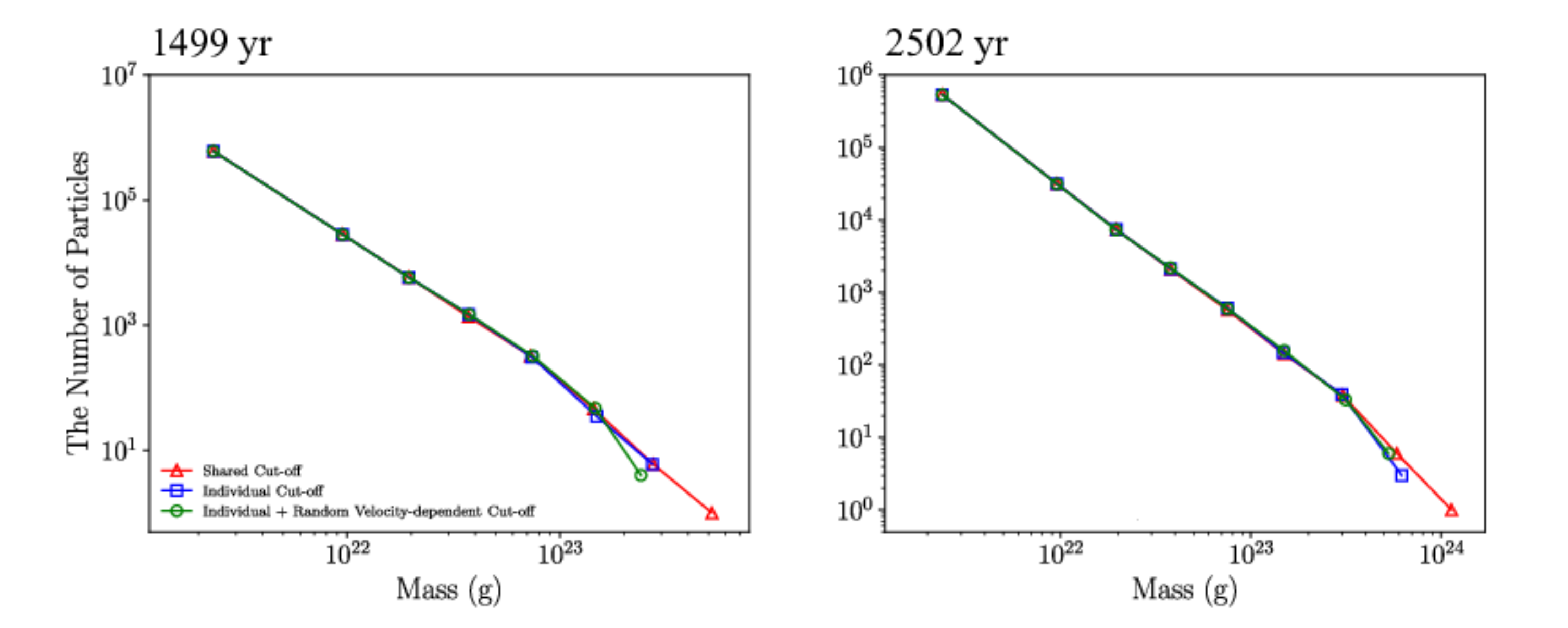}
\end{center}
\caption{
Mass distributions of particles in the case of shared cut-off
(red) and individual cut-off (blue) with perfect accretion at 1499\,yr (left) and 2502\,yr (right). 
We plot the distribution at 2435\,yr in the case of shared cut-off
instead of at 2502\,yr
since the simulation was not performed until 2502\,yr in that case.
}
\label{fig:cum_shaind}
\end{figure*}

\begin{figure*}[htbp]
\begin{center}
\includegraphics[width=0.8\linewidth, bb=0 0 896 663]{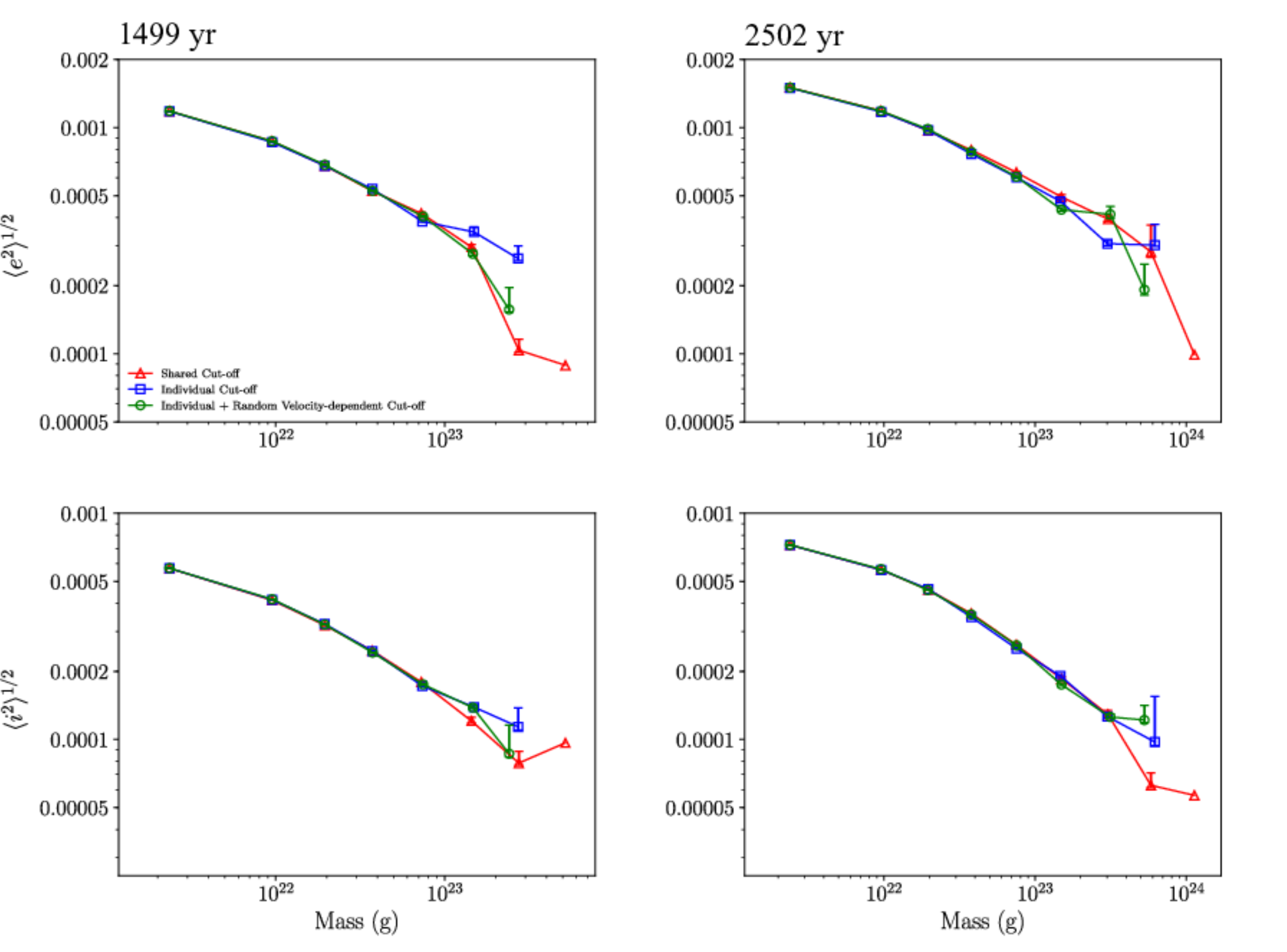}
\end{center}
\caption{
Distribution of rms of orbital eccentricities (top) and inclinations (bottom) of particles as a function of mass in the case of shared cut-off (red), individual cut-off (blue), and individual and random velocity-dependent cut-off with perfect accretion (green) with perfect accretion at 1499\,yr (left) and 2502\,yr (right).
We plot the distribution at 2435\,yr in the case of shared cut-off instead of 2502\,yr 
since the simulation was not performed until 2502\,yr in that case.
The error bars show 70\% confidence intervals.
}
\label{fig:eccinc_long}
\end{figure*}

Figures \ref{fig:cum_shaind} and \ref{fig:eccinc_long} 
 shows the mass distributions and rms random velocities of particles
 at years 1499 and 2502. 
The result does not depend on the choice of the cut-off scheme. Thus,
 we can conclude that the choice of the cut-off scheme does not affect
 the dynamics of the system.


\begin{figure*}[htbp]
\begin{center}
\includegraphics[width=0.45\linewidth, bb=0 0 846 594]{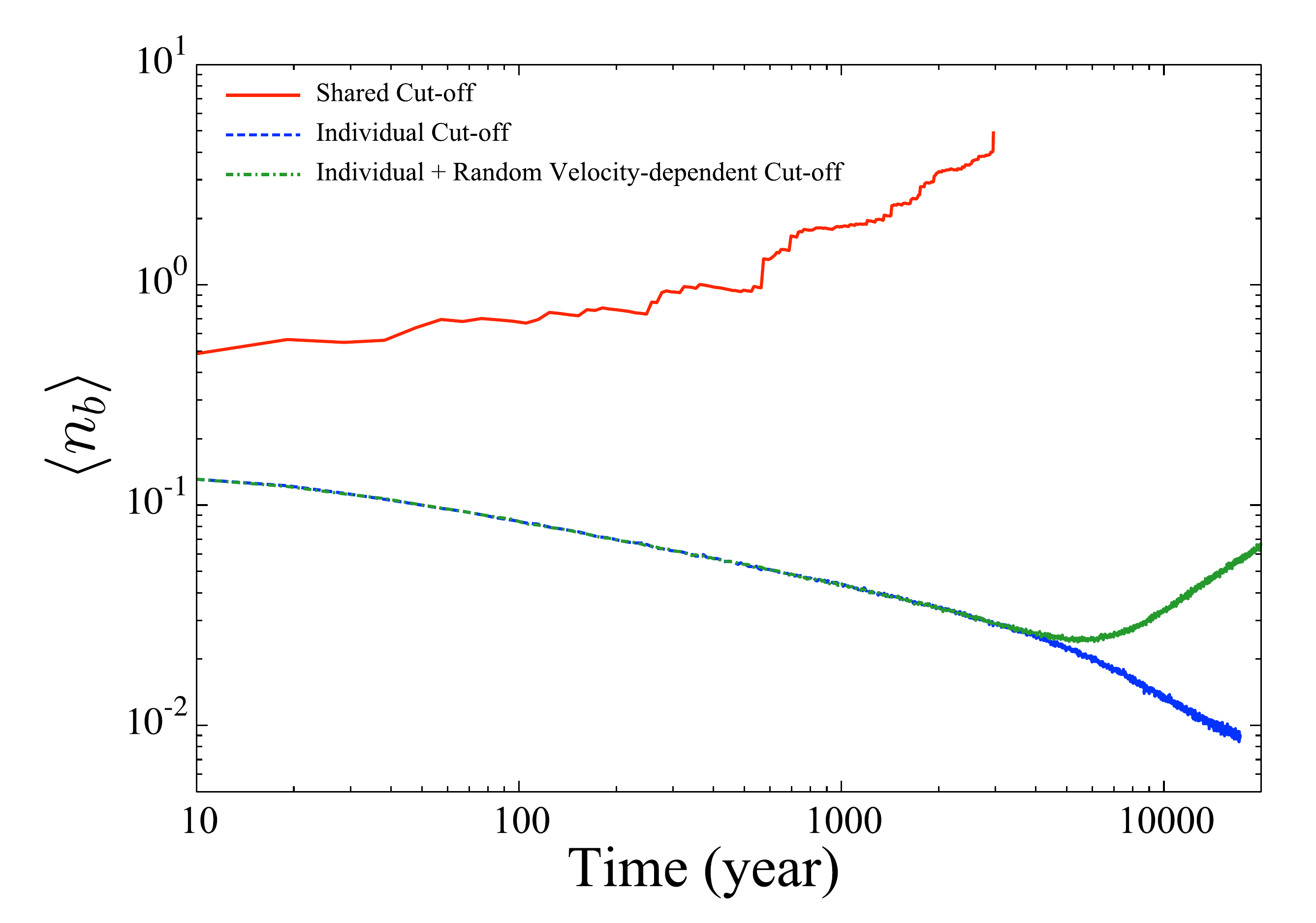}
\includegraphics[width=0.45\linewidth, bb=0 0 846 594]{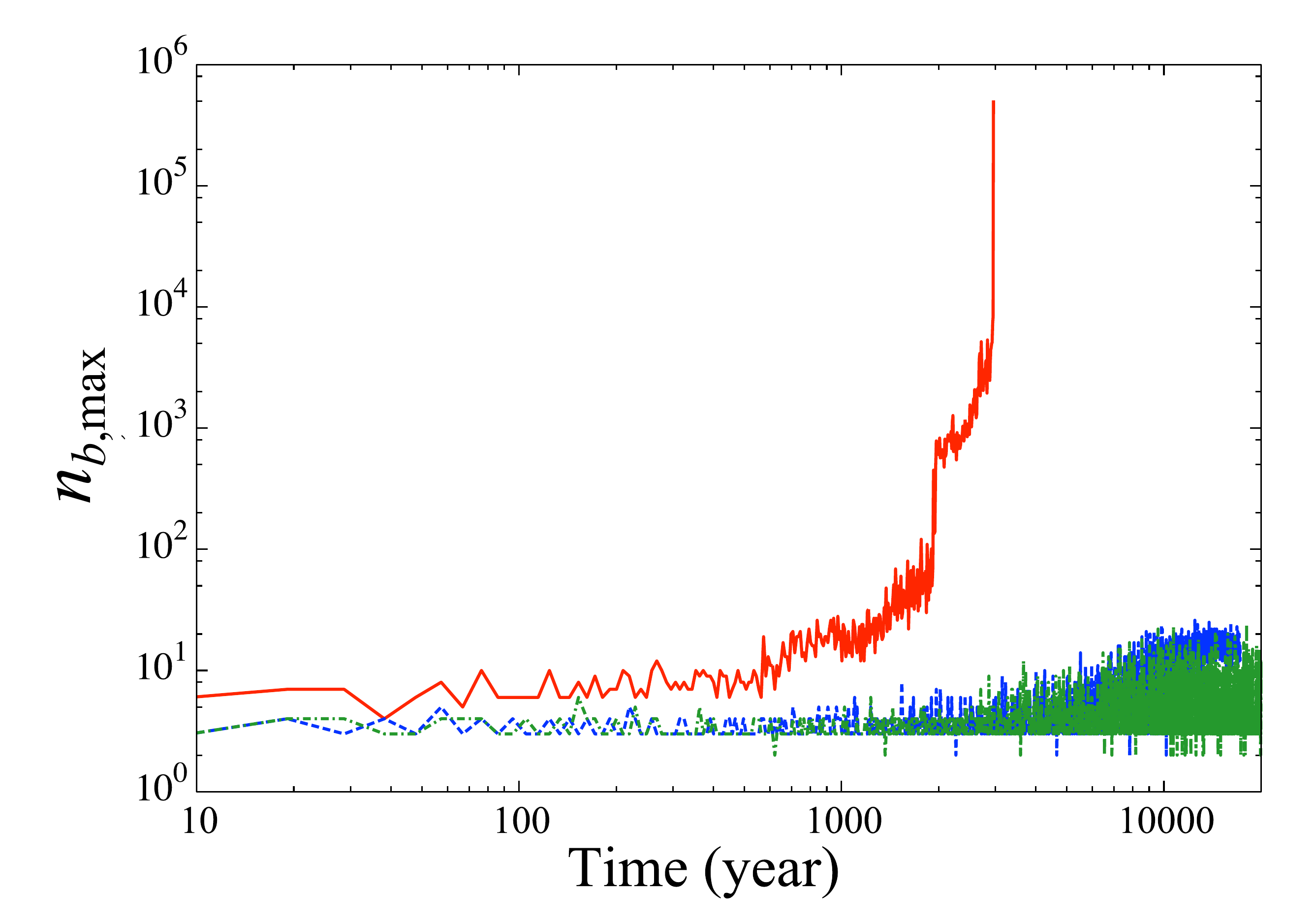}
\end{center}
\caption{
Evolution of the average number of neighbors for each particle, $\langle n_b\rangle$, (left) and the number of particles in the largest neighbor cluster, $n_{b,{\rm max}}$, (right) in the case of shared cut-off (solid red), individual cut-off (dashed blue), and individual and random velocity-dependent cut-off with perfect accretion (dashed-dotted green).
}
\label{fig:clusters}
\end{figure*}

Figure \ref{fig:clusters} shows the average number of neighbors, $\langle n_b\rangle$, and the number
of particles in the largest neighbor cluster, $n_{b,\rm max}$. 
Because the mass of the largest body already reaches about nine times the initial mass at $10\, {\rm yr}$, the average number of neighbors in the case of shared cut-off is larger than in the case of individual cut-off from the beginning of the simulation.
The average number of neighbors, $\langle n_b \rangle$, for the shared cut-off scheme increases
with time, since the shared cut-off radius is determined by the mass of
the most massive particle. On the other hand, that for individual
cut-off, with and without the random velocity term, initially
decreases partly because the total number of particles decreases due to
collisions, and partly because of the increase in the inclination of
particles. 
However, after around $5000\, {\rm yr}$, $\langle n_b\rangle$ for the scheme
with random velocity term starts to increase due to the increase in the
random velocity. 
This increase does not result in a notable increase in the calculation time as can be seen in Fig.
\ref{fig:wtime}. This is simply because $\langle n_b \rangle$  is still very small.

In the case of the shared cut-off scheme, $n_{b,\rm max}$ approached the total number of particles when the calculation was halted. 
This increase in the size of the cluster is of course the reason why the
calculation became very slow. This means that the neighbor cluster
showed percolation, which is expected to occur if $\langle n_b\rangle$ is larger
than the critical value of order unity. When percolation of the 
neighbor cluster occurs, our current implementation falls back to the 
$O(N^2)$ direct Hermite scheme on a single MPI process. Thus, it is
necessary to avoid percolation, and that means we should keep
$\langle n_b\rangle \ll 1$.

\section{Discussion and conclusion\label{s:Conclusion}}

We have presented the implementation and performance of
\ourcode, a parallel $N$-body simulation code based on the $\rm P^3T$
scheme. The main difference from the previous implementation of the 
parallel $\rm P^3T$ scheme \citep{Iwasawa2017} is that we introduced
an individual cut-off radius which depends both on the particle mass  and
the local velocity dispersion. The dependence on the mass is necessary
to handle systems a with wide range of mass spectrum, and 
the local velocity dispersion dependence is necessary to mantain accuracy when the
velocity dispersion becomes high.  With this new treatment of the
cut-off radius, \ourcode can follow a planetary formation process
in which the masses of the  planetesimals  grow by many orders of
magnitude without a significant increase in the calculation time.

We have confirmed that the use of the individual cut-off has no
effect on the result, and that accuracy is improved and the
calculation time is shortened compared to the shared cut-off scheme.

The parallel performance of \ourcode is reasonable for up to 1000
cores. On the other hand, there are systems with much larger numbers of
cores. In particular, the Fugaku supercumputer, which is currently the
fastest computer in the world, has around eight million cores. In order to make
efficient use of such machines, the scalability of \ourcode should be
further improved.

Due to both the distribution of calculation and the increase of communication due to parallelization, there are optimum values for the numbers of parallel MPI and OpenMP.
It should be noted that the optimum values differs depending on the system.

As discussed in section \ref{sect:eqmass}, currently the limiting
factor for the parallel performance is the time for LET
construction, which can be  reduced by several
methods \citep{Iwasawaetal2019}. We plan to apply such methods and
improve the parallel performance.

\ourcode is freely available for all those who are interested in particle simulations. 
The source code is hosted on the GitHub platform and can be downloaded from their site;\footnote{https://github.com/YotaIshigaki/GPLUM}
it has the MIT license.

\section*{Acknowledgements}
This work was supported by MEXT as “Program for Promoting Researches
on the Supercomputer Fugaku” (Toward a unified view of the universe:
from large scale structures to planets).  
This work uses HPCI shared computational resources hp190060 and hp180183.
The simulations in this paper
were carried out on a Cray XC50 system at the Centre for Computational
Astrophysics (CfCA) of the National Astronomical Observatory of Japan
(NAOJ) and a Cray XC40 system at the Academic Center for Computing and Media
Studies (ACCMS) of Kyoto University.  Test simulations were also
carried out on Shoubu ZettaScaler-1.6 at the Institute of Physical and
Chemical Research (RIKEN).  We acknowledge the contribution of
Akihisa Yamakawa, who developed an early version of the parallel $\rm
P^3T$ code.

\end{document}